\documentclass[epj]{svjour}
\usepackage{graphics}
\usepackage{graphicx}
\usepackage{subfigure}
\usepackage{amssymb,amsmath}
\usepackage{CJK}
\usepackage{indentfirst}
\usepackage{amsmath}
\usepackage[square, comma, sort&compress, numbers]{natbib}
\begin{document}
\title{ $Q-\Phi$ criticality in the extended phase space of $(n+1)$-dimensional RN-AdS black holes}
\author{Yu-Bo Ma\inst{1,2},  Ren Zhao\inst{2}, Shuo Cao\inst{1}
\thanks{\emph{e-mail:} caoshuo@bnu.edu.cn}%
}                     
%
%
\institute{ Department of Astronomy, Beijing Normal University,
Beijing, 100875, China \and School of Physic, Shanxi Datong
University, Datong, 037009, China}
%
%
\abstract{ In order to achieve a deeper understanding of gravity
theories, it is important to further investigate the thermodynamic
properties of black hole at the critical point, besides the phase
transition and critical behaviors. In this paper, by using Maxwell's
equal area law, we choose $T,Q,\Phi$ as the state parameters and
study the phase equilibrium problem of general $(n+1)$-dimensional
RN-AdS black holes thermodynamic system. The boundary of the
two-phase coexistence region and its isotherm and isopotential lines
are presented, which may provide theoretical foundation for studying
the phase transition and phase structure of black hole systems.
} 
\maketitle
\section{Introduction}\label{sec:Intro}

In the past analysis, the cosmological constant in Ads space-time
and the state parameter-pressure in general thermodynamic system are
always parallelized as
\begin{equation}
\Lambda=-\frac{n(n-1)}{2 l^2}, P=\frac{n(n-1)}{16 \pi l^2}
\end{equation}
and the corresponding thermodynamics volume for black hole
thermodynamic system expresses as
\begin{equation}
 V=(\frac{\partial M}{\partial P})_{S,Q_{i},J_{k}}
\end{equation}
where T,P, and V are the variable state parameters.
\cite{3,4,5,6,7,8,13,16,17,18,19,21,22,23,24,25,26,30,31,34,35,36,37,40,47,48}.
 This traditional method has been extensively applied to the characterization of
general thermodynamic systems, and also to the establishing of
complete simulation for Ads Space-time black hole thermodynamic
liquid /gas system
\cite{1,2,9,10,11,12,14,15,20,27,28,29,32,33,38,39,41,42,43}.

A more recent method is to choose state parameters $(T,P,V)$ as
variables and apply the Ehrenfest scheme to the study of various
black hole critical phenomenon in Ads space-time. It has been proved
that at the phase transition point, the second-order phase
transition exists in Ads Space-time black holes and the
thermodynamics agree very well with Ehrenfest equation
\cite{20,21,24,26,27,44,45}. At present there are two common methods
to investigate the the critical phenomenon of Ads Space-time Black
Holes in the literature. The first is study the thermodynamics and
state space geometry of black holes
\cite{28,29,30,31,32,42,46,47,48,49,57}, which found that the black hole
phase transition point meets the requirements of thermodynamic
second-order phase transition. The second approach is to turn to
Maxwell's equal area law and discuss the critical behavior of Ads
Space-time black hole system \cite{31,50,51,52,53,54}, which have
also proved the existence of second-order phase transition at Black
Hole phase transformation point.

Despite of the promising results obtained about thermodynamic
properties of Ads Space-time black holes, from theoretical point of
view, there should be critical behaviors and phase transition
process, if taking black holes in Ads Space-time as thermodynamic
systems. However, the statistical mechanics background of Black
Holes as thermodynamic systems is still unknown, which makes it very
meaningful to study the relations between different thermodynamic
properties of Ads Space-time black holes. Moreover, such study will
contribute to a deeper understanding of the thermodynamic properties
of black hole (entropy, temperature, and heat capacity), as well as
the completion of self-consistent geometry theory of black hole
thermodynamics.

The $P-V$ phase diagram of Ads Space-time black holes was analyzed
in Ref.~\cite{10}, which implied the existence of a mechanical
unstable region when the black hole temperature is low. In this
region with partial negative pressure, pressure increases together
with volume ${\partial M}/{\partial P}>0$ in the isotherm, which is
similar to the result obtained from the $ P-V$ phase diagram of van
der Waals-Maxwell liquid /gas. A possible solution to the famous
Maxwell's equal area law, which was extensively applied to the study
of Ads Space-time black holes thermodynamic system
\cite{49,50,51,52,53,54}. The yielded $T-P$ curve for system biphase
equilibrium and the slope expression of biphase equilibrium curve
showed that Ads Space-time black hole has the second-order phase
transition and can be in a two phase coexistence state. We remark
here that, except the phase transition point, other phase
transitions in Ads Space-time black hole are all of first order.
However, all the above results were derived on the base of the
precondition of invariance of electric charge. As is well known to
everyone, the parameters describing the charged Ads Space-time Black
Holes thermodynamic system are not only related to state parameters
$(T,P,V)$ , but also electromagnetic parameters like charge and
electric potential. In this paper, we will use Maxwell's equal area
law to study the thermodynamic properties of general
$(n+1)$-dimensional RN-AdS black holes. More specifically, the
discussion of the following two problems is the main motivation of
our analysis: For black hole thermodynamic system with constant $P$
and $V$, is there still second-order phase transition when taking
$(T,Q,\Phi)$ as state parameters? If so, is the critical point still
the same as that when $(T,P,V)$ are taken as state parameters? In
the second part, we give a brief introduction of $(n+1)$-dimensional
RN-AdS black hole. In the third part, we apply Maxwell's equal area
law in $(n+1)$-dimensional RN-AdS black hole thermodynamic system,
and obtain the relationship between different parameters and the
boundary of two phase coexistence region. The last part is the
conclusion.

\section{ general $(n+1)$-dimensional RN-AdS black holes}
\label{sec:1}

For general $(n+1)$-dimensional RN-AdS black holes, the space-time
metric can be written as \cite{32}
\begin{equation}
d{{s}^{2}}=-f(r)d{{t}^{2}}+\frac{d{{r}^{2}}}{f(r)}+{{r}^{2}}d\Omega _{n-1}^{2} ,
\end{equation}
where $d\Omega_{n-1}$ is the metric of the associated
$(n+1)$ dimensional base manifold and
\begin{equation}
f(r)=k-\frac{8\Gamma (\frac{n}{2})M}{(n-1){{\pi }^{\frac{n}{2}-1}}{{r}^{n-2}}}+\frac{{{Q}^{2}}}{{{r}^{2n-4}}}+\frac{{{r}^{2}}}{{{l}^{2}}} ,
\end{equation}
Here $k=1,0,-1$ respectively corresponds to the sphere, plane and
hyperbola symmetric cases. If denoting $r_+$ as the position of
black hole horizon satisfying $f(r_+)=0$, one can straightforwardly
obtain the value of $r_+$ , with the mass of the black hole within
the event horizon radius
\begin{equation}
M=\frac{(n-1){\pi^{\frac{n}{2}-1}}{r_+}^{-n-2}({{l}^{2}}{{Q}^{2}}{{r}_{+}}^{4}+{{l}^{2}}{{r}_{+}}^{2n}k+{{r}_{+}}^{2n+2})}{8{{l}^{2}}\Gamma(\frac{{n}}{2})} ,
\end{equation}
Correspondingly, the Hawking temperature, entropy and potential of
the black hole could be obtained as
\begin{equation}
 T=\frac{{f}'({{r}_{+}})}{4\pi }=\frac{1}{4\pi }(\frac{k(n-2)}{r_+}-(n-2){{Q}^{2}}{{r_+}^{3-2n}}-\frac{2r_+\Lambda }{n-1}),
\end{equation}
\begin{equation}
  S=\int\limits_{0}^{{{r}_{+}}}{\frac{{{\partial }_{{{r}_{+}}}}M\left( {{r}_{+}},Q \right)}{T}=\frac{{{\pi }^{\frac{n}{2}}}{{r}_{+}}^{n-1}}{2\Gamma \left( \frac{{n}}{2} \right)}},
\end{equation}
\begin{equation}
  \Phi ={{\left( \frac{\partial M}{\partial Q} \right)}_{S}}=\frac{\left( n-1 \right){{\pi }^{\frac{n}{2}-1}}Q{{r}_{+}}^{2-n}}{4\Gamma \left( \frac{n}{2} \right)},
\end{equation}
When taking $k=1$, from Eq.~(6), we will get
\begin{equation}
 {{Q}^{\frac{1}{n-2}}}=-\frac{n-1}{4\Lambda }A\left( 4\pi T-\sqrt{16{{\pi }^{2}}{{T}^{2}}-\frac{8\Lambda (n-2)}{n-1}\left( {{A}^{2n-4}}-1 \right)}\right),
\end{equation}
where $A={{\left( \frac{4\Gamma (n/2)}{(n-1){{\pi }^{n/2-1}}}\Phi
\right)}^{1/(n-2)}}$. From the derivative of Eq.~(9)
\begin{equation}
{{\left( \frac{\partial Q}{\partial \Phi } \right)}_{T}}=0,
\end{equation}

\begin{equation}
{{\left( \frac{{{\partial }^{2}}Q}{\partial {{\Phi }^{2}}} \right)}_{T}}=0,
\end{equation}
we will obtain the following expression
\begin{equation}
(n-2)(2n-3){{A}^{2n-2}}-(n-2){{A}^{2}}-\frac{2\Lambda }{n-1}{{Q}^{2/(n-2)}}=0,
\end{equation}
\begin{equation}
(n-1)(2n-3){{A}^{2n-4}}-1=0,
\end{equation}
Combining Eqs.~(9),(12) and (13), one can obtain the three quantities
under critical condition
\begin{equation}
{{\Phi }_{c}}=\frac{{{\pi }^{n/2-1}}}{4\Gamma (n/2)}\sqrt{\frac{(n-1)}{(2n-3)}},
\end{equation}
\begin{equation}
{{T}_{c}}=\frac{(n-2)}{\pi (2n-3)}{{(-2\Lambda )}^{1/2}},
\end{equation}
\begin{equation}
{{Q}_{c}}={{\left( -\frac{{{(n-2)}^{2}}}{2\Lambda } \right)}^{(n-2)/2}}{{\left( (n-1)(2n-3) \right)}^{-1/2}},
\end{equation}

The critical values for $Q$, $\Phi$, $T$ have been given in Eqs.(14)-(16). According to these critical values, the critical ratio is given by
\begin{equation}
{{\rho}_{c}}=\frac{\sqrt{-\Lambda } 2^{-\frac{n}{2}-\frac{3}{2}} (n-1) \pi ^{n/2} \left(-\frac{(n-2)^2}{\Lambda }\right)^{n/2}}{(n-2)^3 \sqrt{\frac{n-1}{2 n-3}} \sqrt{2 n^2-5 n+3} \Gamma \left(\frac{n}{2}\right)},
\end{equation}
Obviously, not like the P-V criticality, it depends on $\Lambda$ and $n$.

And, when taking $n=3,4,5,6$ and $\Lambda=-1$, one can get the
$Q-\Phi$ graphs at different temperature from the above equations,
as well as different critical temperatures under different
space-time dimension, respectively ${{T}_{c3}}=0.150053$,
${{T}_{c4}}=0.180063$, ${{T}_{c5}}=0.192925$ and
${{T}_{c6}}=0.20007$. Fig.~1 shows the $Q-\Phi$ diagram at different
critical temperature $T_c$, from which one can see that $Q-\Phi$
curve intersect with x-axis at $\Phi=\Phi_c$. It is apparent that
when the temperature $T>T_c$, the special region with ${{\left(
\frac{\partial Q}{\partial \Phi } \right)}_{T}}<0$ indeed exist in
the $Q-\Phi$ diagram, which does not satisfy the requirements of
thermodynamic stability in the process of black hole evolution. And in Fig.2, we also plot the $Q-\Phi$ curves at different values of $\Lambda$. It is shown that the $Q-\Phi$ criticality nearly unchanges. Only the position of the critical point chandes.


\begin{figure*}[htbp]
  \begin{center}
    \mbox{
      \subfigure[$n=3$]{\scalebox{0.35}{\includegraphics{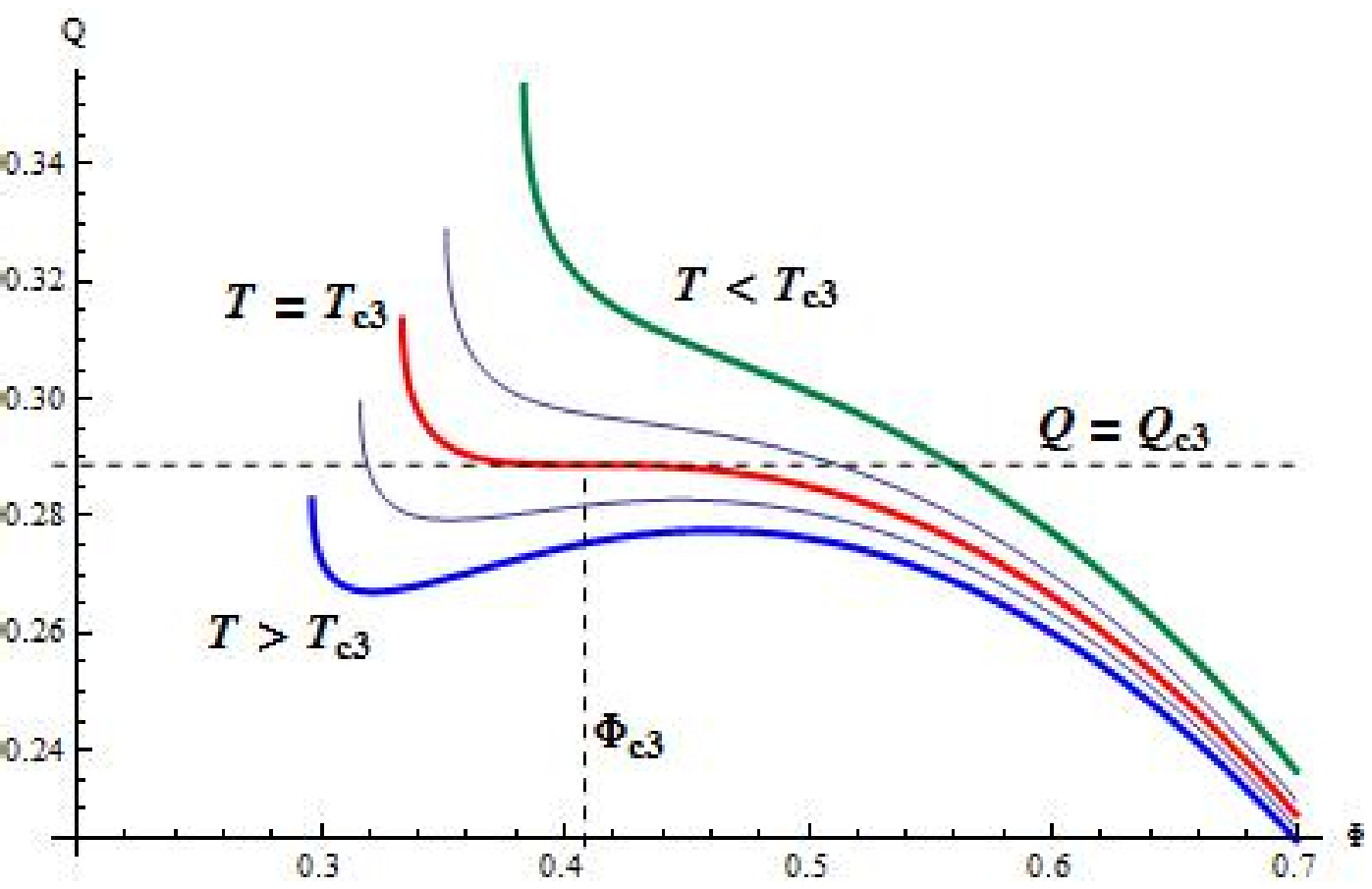}}} \quad
      \subfigure[$n=4$]{\scalebox{0.35}{\includegraphics{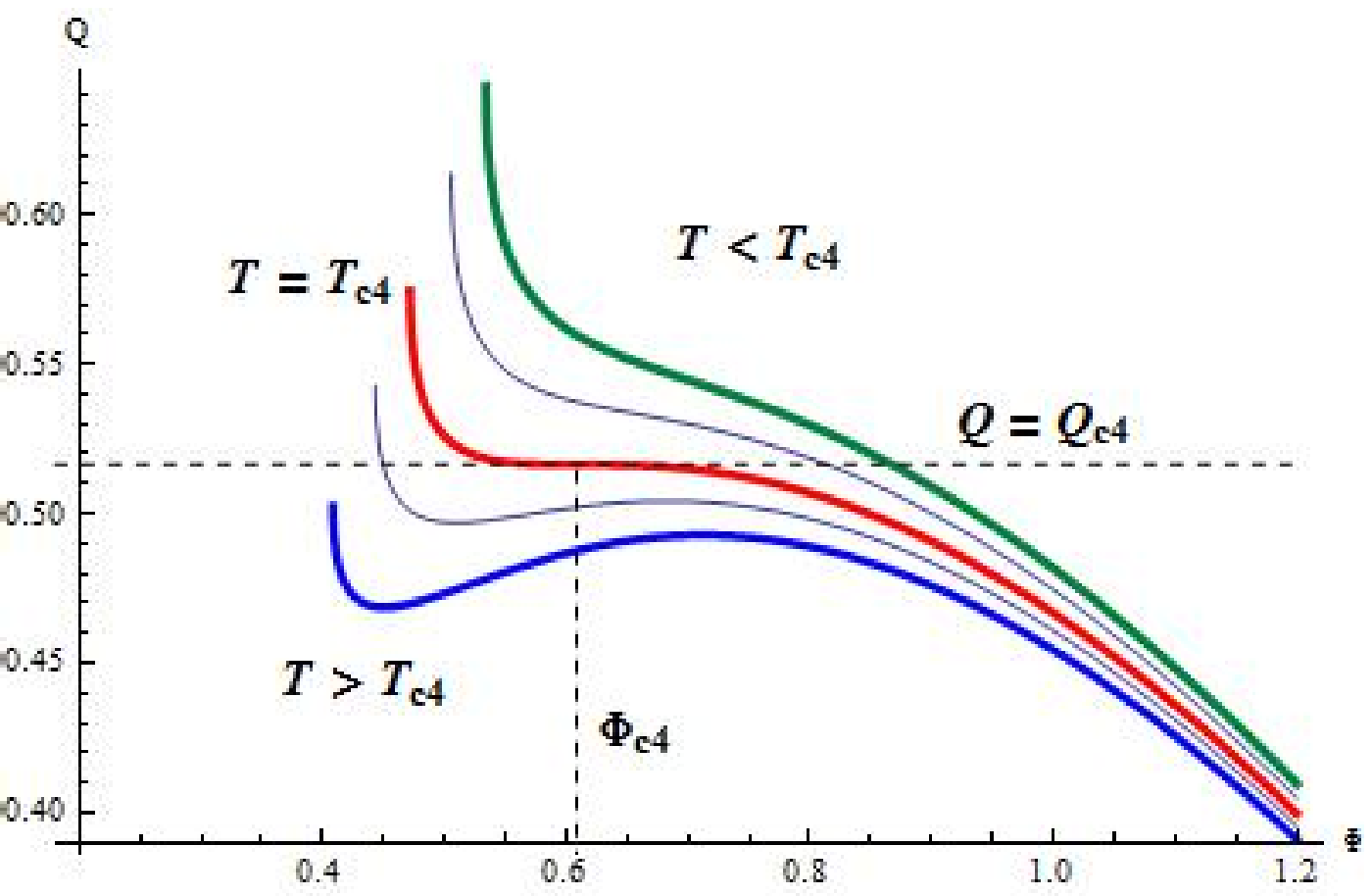}}}
      }
    \mbox{
      \subfigure[$n=5$]{\scalebox{0.35}{\includegraphics{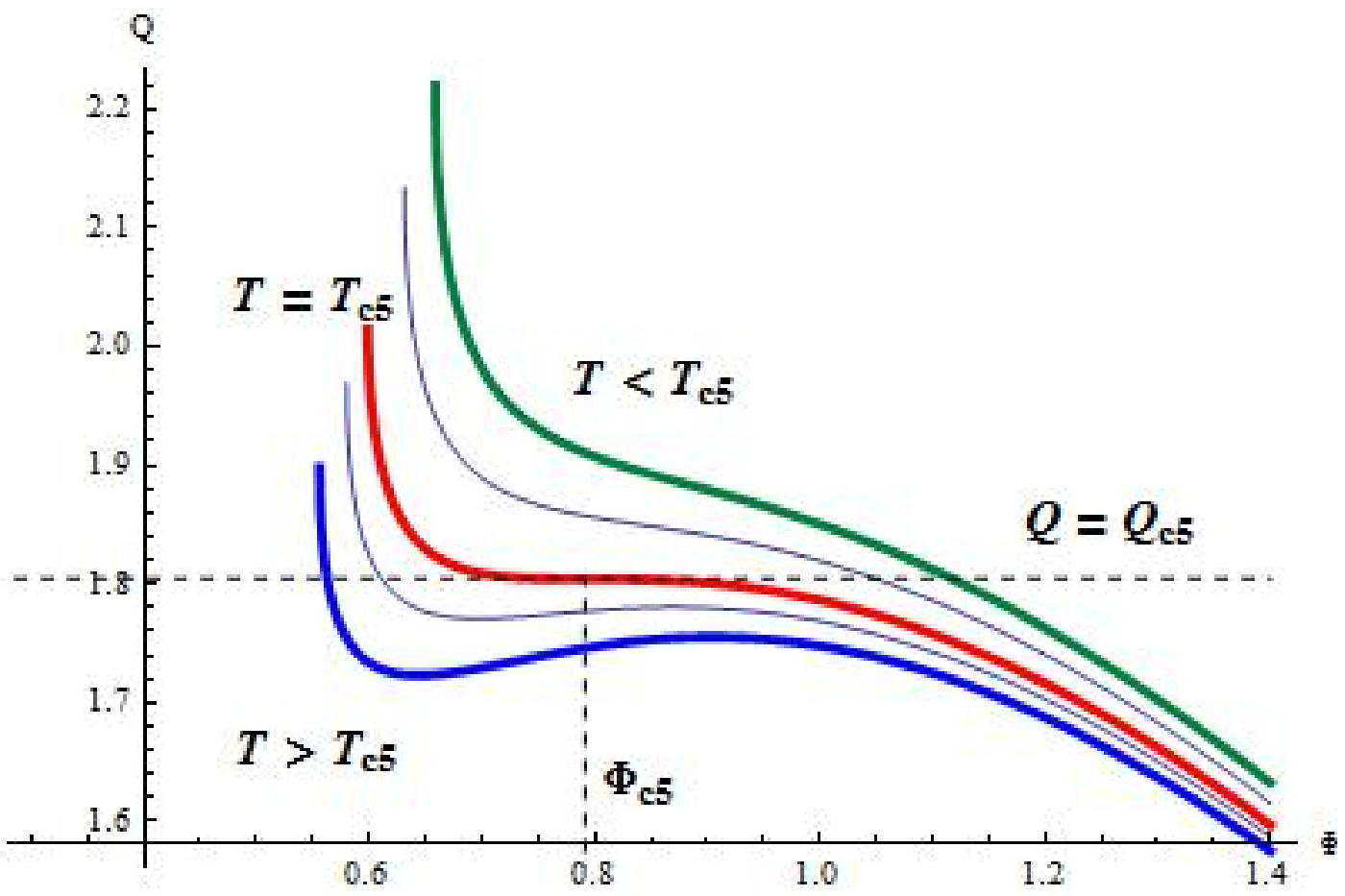}}} \quad
      \subfigure[$n=6$]{\scalebox{0.35}{\includegraphics{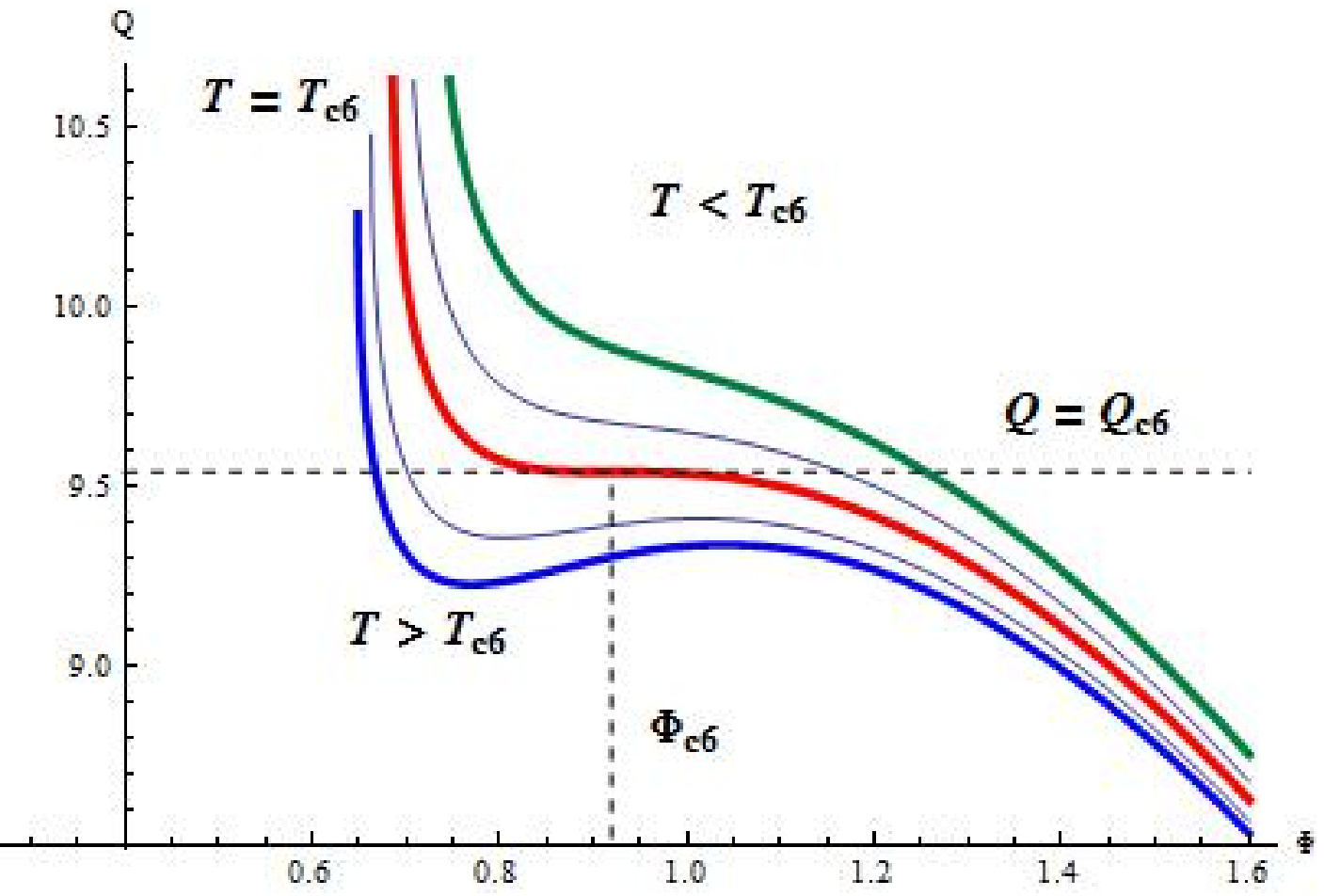}}}
      }
    \caption{$Q-\Phi$ diagram under isothermal conditions when the parameter $n$ is fixed at $n=3,4,5,6$, respectively.
    }
    \label{Qn}
  \end{center}
\end{figure*}

\begin{figure*}[htbp]
  \begin{center}
    \mbox{
      \subfigure[$n=3,\Lambda=-0.5$]{\scalebox{0.5}{\includegraphics{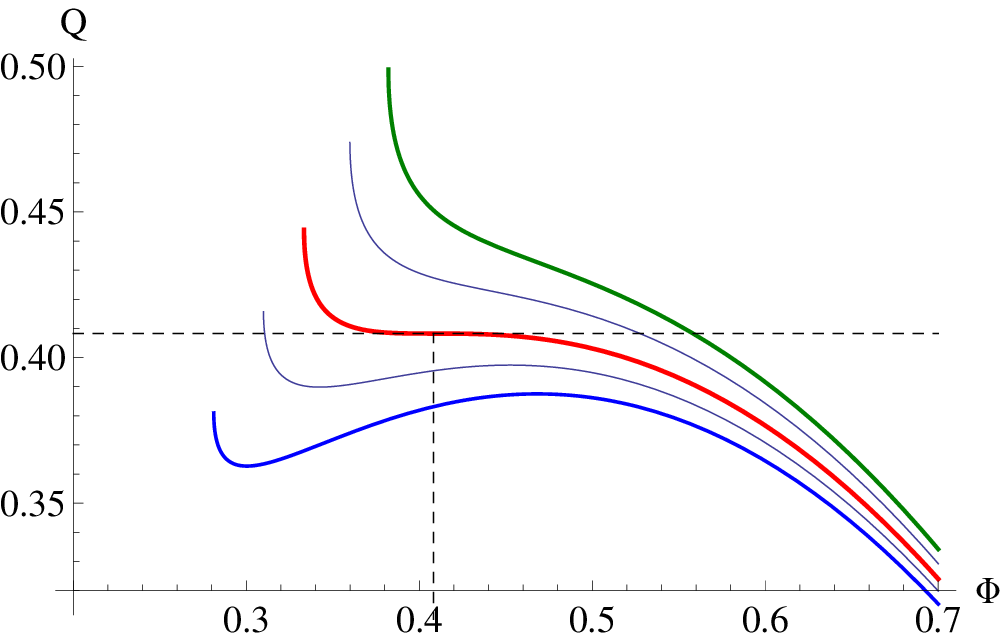}}} \quad
      \subfigure[$n=3,\Lambda=-1$]{\scalebox{0.5}{\includegraphics{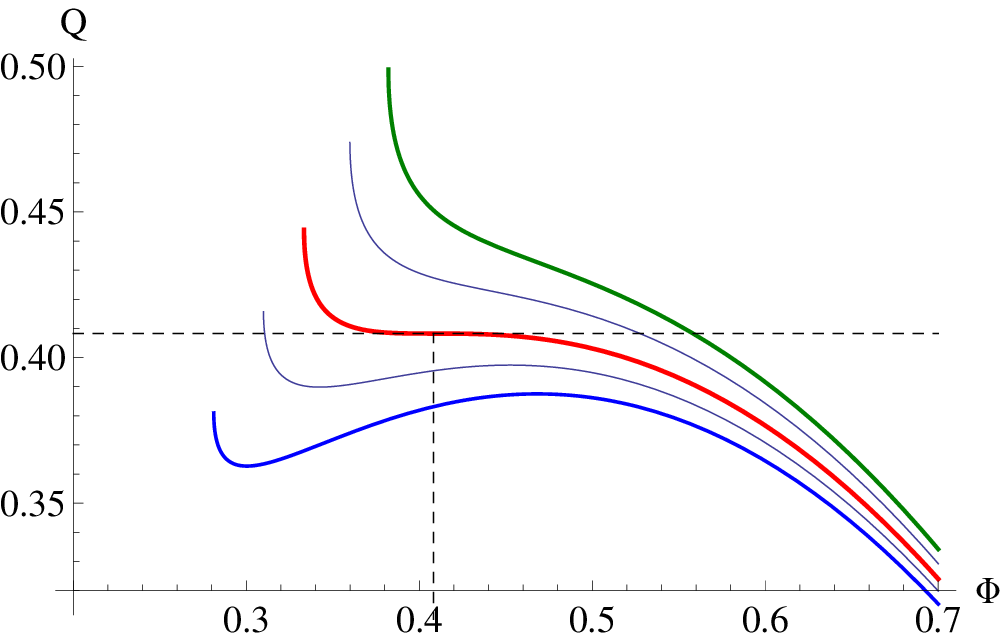}}}
      }
    \mbox{
      \subfigure[$n=3,\Lambda=-2$]{\scalebox{0.5}{\includegraphics{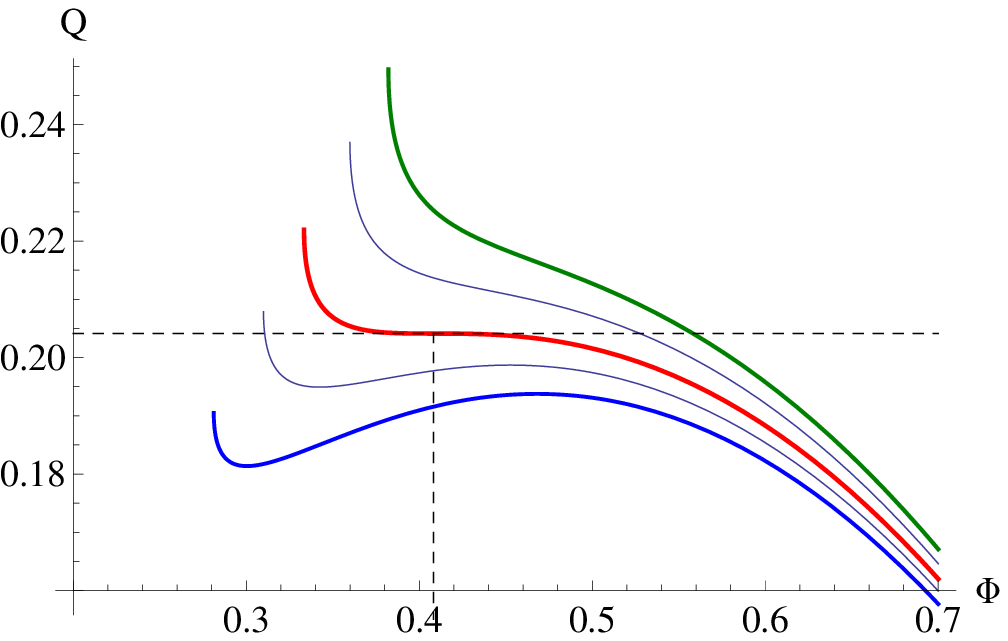}}} \quad
      \subfigure[$n=3,\Lambda=-3$]{\scalebox{0.5}{\includegraphics{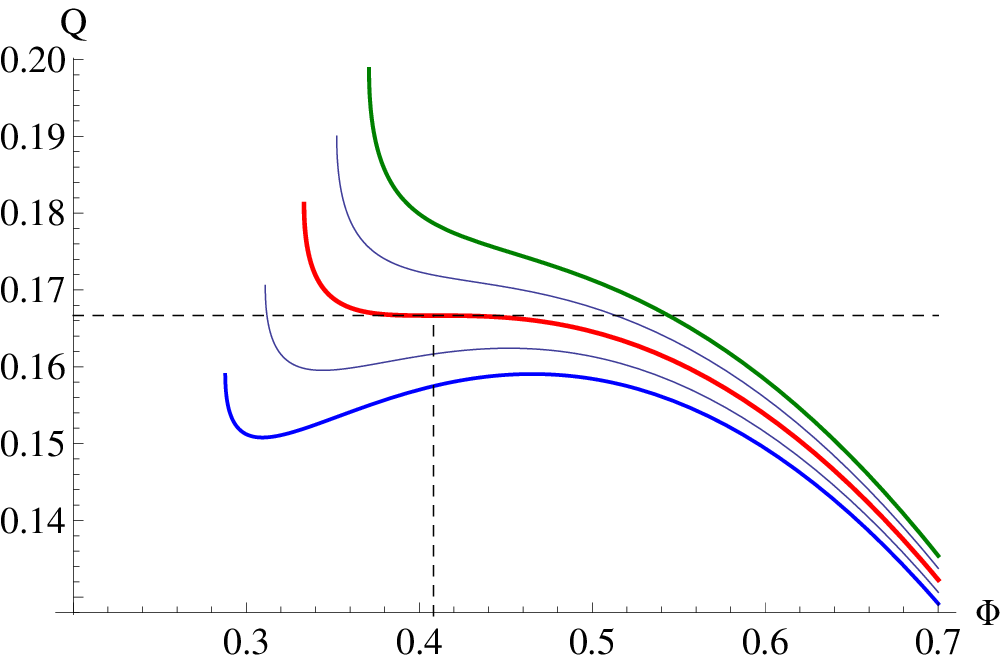}}}
      }
    \caption{$Q-\Phi$ diagram under isothermal conditions when the parameter $n=3$ is fixed and $\Lambda=-0.5,-1,-2,-3$, respectively.
    }
    \label{Qn}
  \end{center}
\end{figure*}

Another important thermodynamical quantity is the heat capacity
$C_Q$ at constant charge, which measures the stability against small
perturbation and could be defined as
\begin{equation}
\begin{aligned}
 {{C}_{Q}} & = T{{\left( \frac{{{\partial }}S}{{{\partial }}T} \right)}_{Q}} \\
           & = \frac{-\left( 2{{\left( -1+n \right)}^{2}}{{\pi }^{1+\frac{n}{2}}}{{Q}^{\frac{1}{-2+n}}}{{r_+}^{-1+n}}{{\left( Q{{r_+}^{2-n}} \right)}^{\frac{3}{-2+n}}} \right)}{\Delta }T,
\end{aligned}
\end{equation}
where
\begin{equation}
\begin{aligned}
\Delta = & \Gamma (\frac{n}{2})(-2{{n}^{3}}{{\left( {{\left( Q{{r_+}^{2-n}} \right)}^{\frac{1}{-2+n}}} \right)}^{2n}}\\
         & +{{n}^{2}}\left( {{\left( Q{{r_+}^{2-n}} \right)}^{\frac{4}{-2+n}}}+9{{\left( {{\left( Q{{r_+}^{2-n}} \right)}^{\frac{1}{-2+n}}} \right)}^{2n}} \right) \\
         & -n\Gamma (\frac{n}{2})\left( 3{{\left( Q{{r_+}^{2-n}} \right)}^{\frac{4}{-2+n}}}+13{{\left( {{\left( Q{{r_+}^{2-n}} \right)}^{\frac{1}{-2+n}}} \right)}^{2n}} \right) \\
         & +2\Gamma (\frac{n}{2})\left( {{\left( Q{{r_+}^{2-n}} \right)}^{\frac{4}{-2+n}}}+3{{\left( {{\left( Q{{r_+}^{2-n}} \right)}^{\frac{1}{-2+n}}} \right)}^{2n}} \right)\\
         & +2\Gamma (\frac{n}{2}){{Q}^{\frac{2}{-2+n}}}{{\left( Q{{r_+}^{2-n}} \right)}^{\frac{2}{-2+n}}}\Lambda,
\end{aligned}
\end{equation}

For comparison, in Fig.~3 we show the heat capacity $C_Q$ changing
with ${{r}_{+}}$ and $Q$, fixing $\Lambda=-1$ and $\Lambda=-0.5$ at
$n=3$. As can be seen from Fig.~3(a), for small value of $Q$, with
the increase of ${{r}_{+}}$, the heat capacity first goes to
positive infinity at $r_{+}={{r}_{1}}$, then increases from negative
infinity to a finite negative value and goes back to negative
infinity at ${{r}_{+}}={{r}_{2}}$. Finally, it will decrease from
positive infinity to a finite positive value, and then monotonically
increases to infinity at ${{r}_{+}}=\infty $. From the above
analysis, it is quite evident that the heat capacity ${{C}_{Q}}$ is
positive for ${{r}_{+}}<{{r}_{1}}$ and ${{r}_{+}}>{{r}_{2}}$, while
negative for ${{r}_{2}}<{{r}_{+}}<{{r}_{1}}$. Note that ${{r}_{1}}$
represents the transition point where a small stable black hole with
${{C}_{Q}}>0$ changes to an intermediate unstable one with
${{C}_{Q}}<0$. ${{r}_{2}}$ corresponds to the transition point at
which an intermediate unstable black hole changes to a large stable
one.

\begin{figure*}[htbp]
  \begin{center}
    \mbox{
      \subfigure[$\Lambda=-1$]{\scalebox{0.4}{\includegraphics{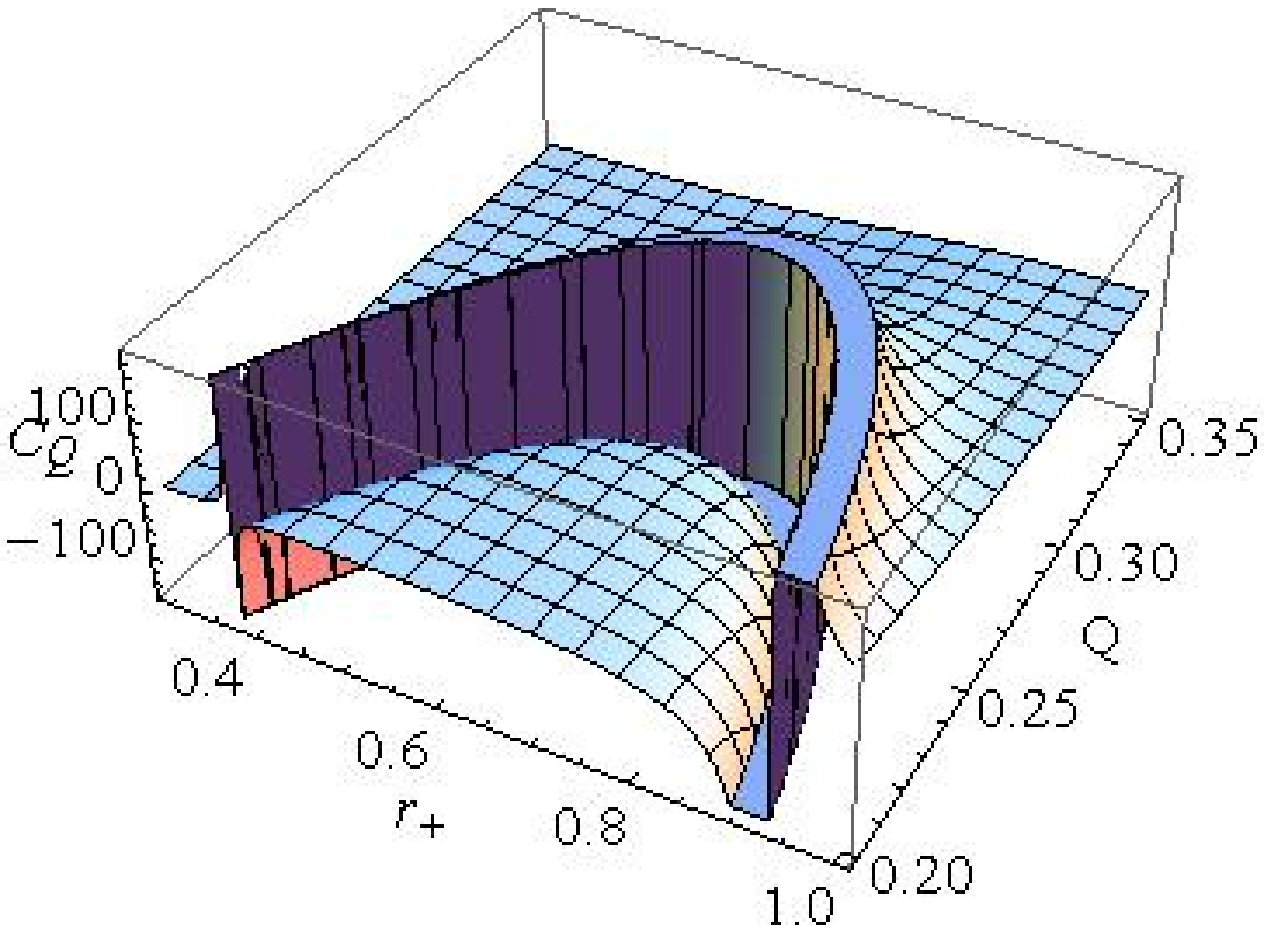}}} \quad
      \subfigure[$\Lambda=-0.5$]{\scalebox{0.4}{\includegraphics{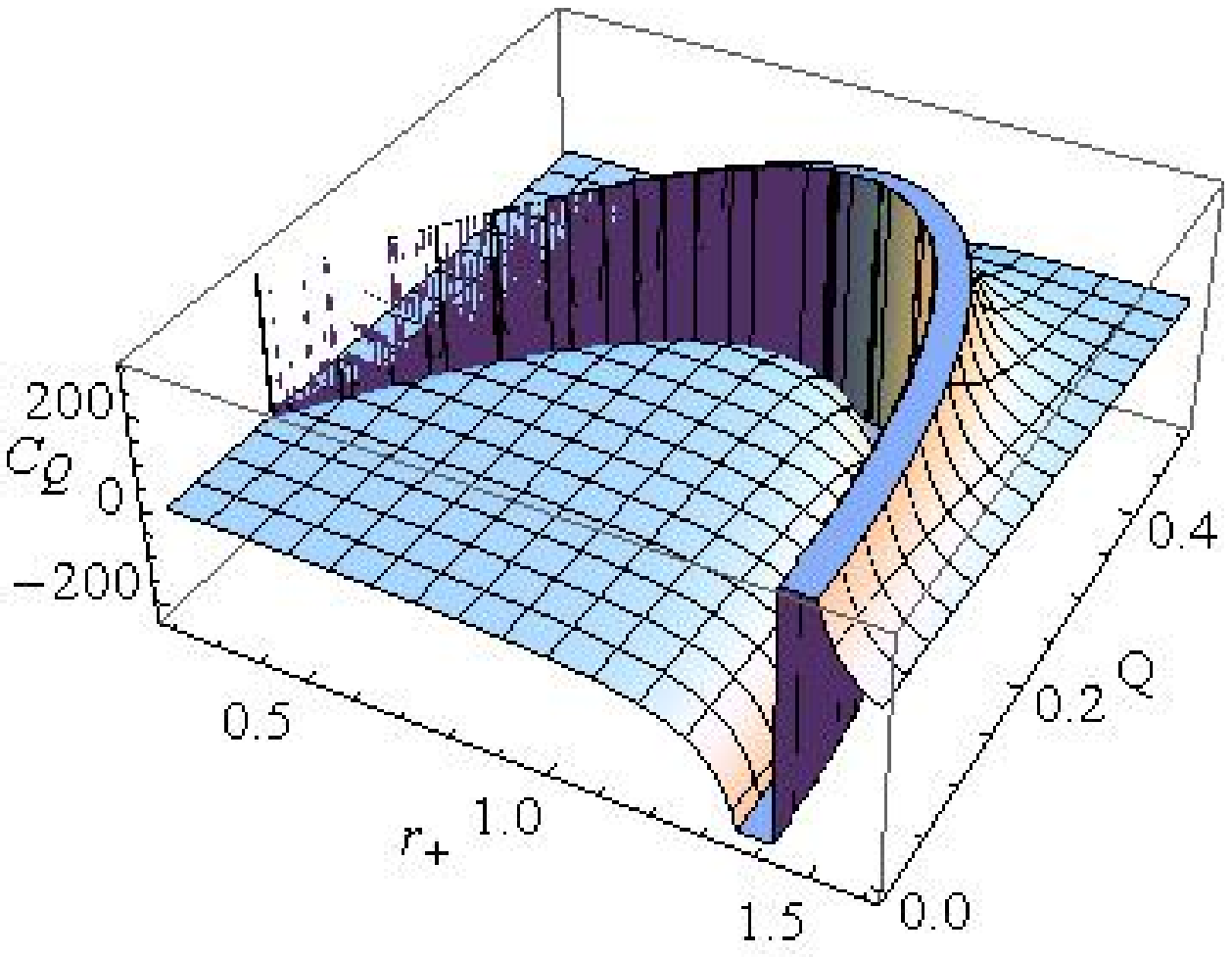}}}
      }
       \caption{Heat capacity $C_Q$ varying with $r_+$ and $Q$. (a) for $n=3, \Lambda=-1$, (b) for $n=3,
       \Lambda=-0.5$.
       }
  \end{center}
\end{figure*}

On the other hand, for large value of $Q$, we find the divergent
behavior vanishes and the black hole will stay in a stable phase. In
order to obtain a better understanding of the divergent behaviors of
the heat capacity ${{C}_{Q}}$, we illustrate in Fig.~4(a) the
divergent point in the ($Q$,${{r}_{+}}$) plane. We emphasize $T<0$
represents a non-black hole case which will not be discussed here.
For $Q<{{Q}_{c3}}$, there exist two divergent points at ${{r}_{1}}$
and ${{r}_{2}}$; for $Q={{Q}_{c3}}$, the two divergent points
coincide with each other at ${{r}_{+}}={{r}_{c3}}$; while the
divergent point disappears at $Q>{{Q}_{c3}}$. Therefore,
${{Q}_{c3}}$ represents a critical phase transition point, which
corresponds to a local maxima along this divergent curve. Moreover,
this critical value also varies with the parameter $\Lambda$, as can
be seen from both Fig.~3(b) and Fig.~4(b).

\begin{figure*}[htbp]
  \begin{center}
    \mbox{
      \subfigure[$\Lambda=-1$]{\scalebox{0.4}{\includegraphics{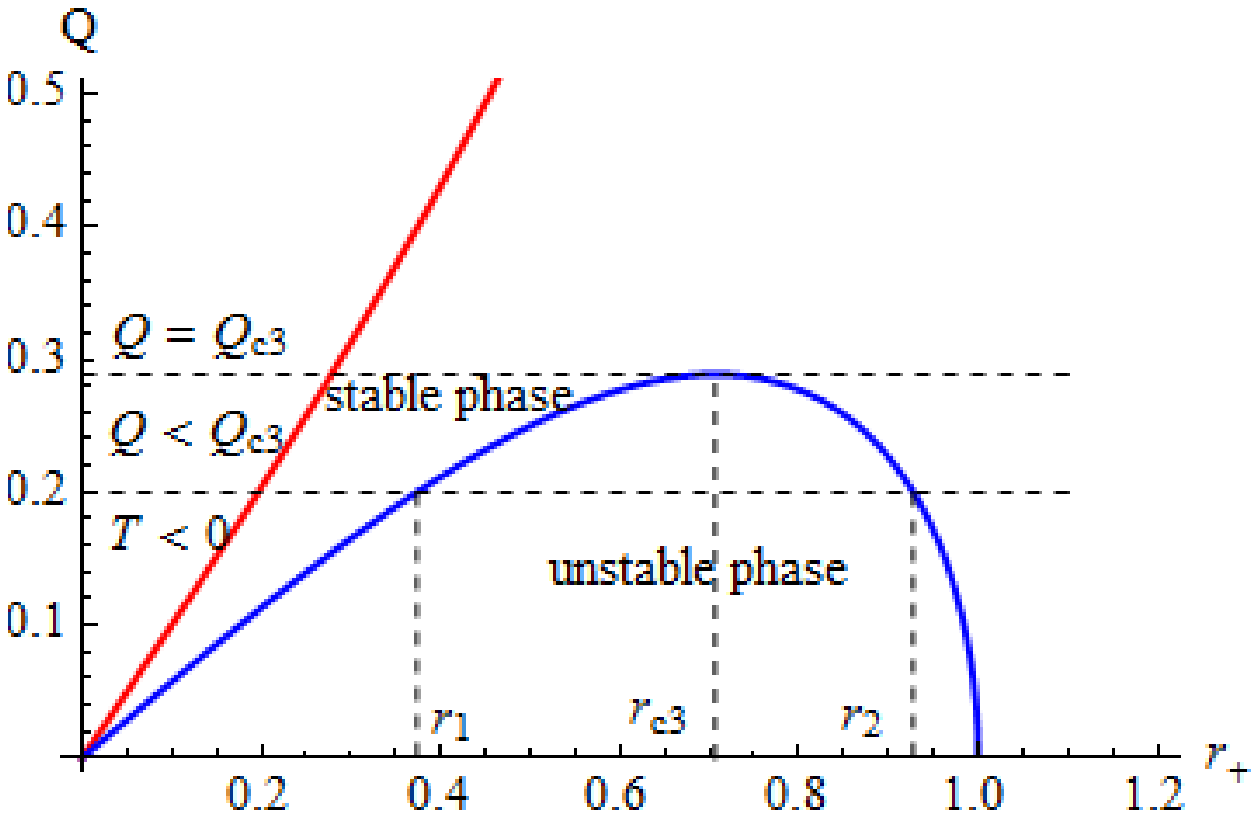}}} \quad
      \subfigure[$\Lambda=-0.5$]{\scalebox{0.4}{\includegraphics{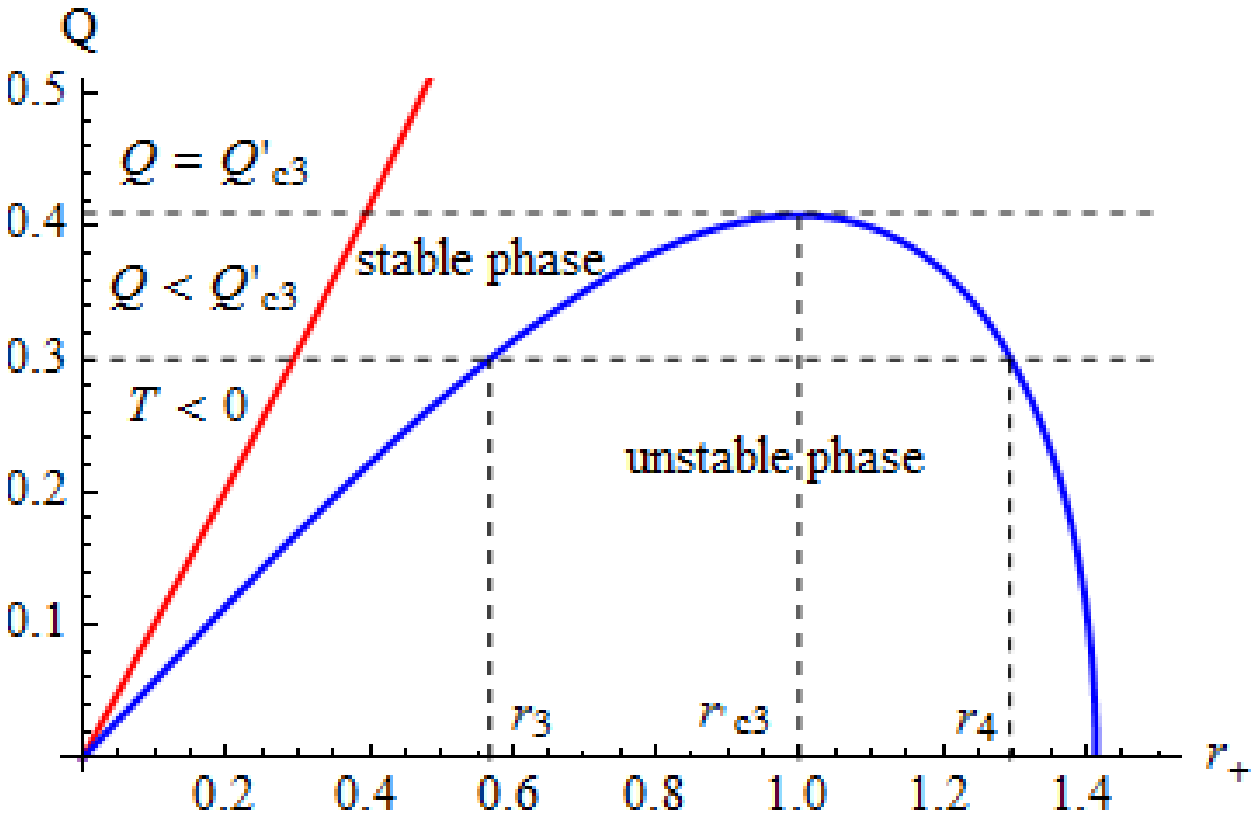}}}
      }
       \caption{Divergent behavior of the heat capacity as a function of $r_+$. (a) for $n=3, \Lambda=-1$, (b) for $n=3,
       \Lambda=-0.5$.
       }
  \end{center}
\end{figure*}

\section{Phase equilibrium and Maxwell's equal area law}\label{sec:2}

For general $(n+1)$-dimensional RN-AdS black holes with constant
temperature, the $Q-\Phi $ curve shows an unstable region with
${{\left( \frac{\partial Q}{\partial \Phi } \right)}_{T}}<0$. This
is similar to Ads Space-time black holes with $(T,P,V)$ as state
parameters. However, these problems were all solved when taking the
gas to liquid phase transition \cite{31,50,53}. In this section, we
will study the famous Maxwell's equal area law in Van der Waals
equation, and apply it to the phase transition of general
$(n+1)$-dimensional RN-AdS black holes. Here we choose $(T,Q,\Phi)$
to be the state parameters and obtain the two phase coexistence
region boundary. In this region, isotherm of general
$(n+1)$-dimensional RN-AdS black holes is replaced by isopotential
line. The maximum of coexistence curve is the so-called critical
point, at which Maxwell's equal area law is no longer applicable.

Assuming the temperature is higher than the critical temperature
$(T_0> T_c)$, we set the x-axis of the boundary of two-phase region
to be $\Phi _{2}^{1/(n-2)}={{\tilde{\Phi }}_{2}}$ and $\Phi
_{1}^{1/(n-2)}={{\tilde{\Phi }}_{1}}$, while the y-axis to be
$Q_{0}^{1/(n-2)}={{\tilde{Q}}_{0}}$. From Maxwell's equal area law,
we have
\begin{equation}
 {{\tilde{Q}}_{0}}({{\tilde{\Phi }}_{2}}-{{\tilde{\Phi }}_{1}})=\int\limits_{{{{\tilde{\Phi }}}_{2}}}^{{{{\tilde{\Phi }}}_{1}}}{\tilde{Q}d\tilde{\Phi }},
\end{equation}
Taking $A=\tilde{A}\tilde{\Phi}$ and $\tilde{A}={{\left(
\frac{4\Gamma (n/2)}{(n-1){{\pi }^{n/2-1}}} \right)}^{1/(n-2)}}$,
Eq.~(9) will transform into
\begin{equation}
\begin{aligned}
  \tilde{Q}=& -\frac{n-1}{4\Lambda }\tilde{A}\tilde{\Phi }\\
            & \left( 4\pi T-\sqrt{16{{\pi }^{2}}{{T}^{2}}-\frac{8\Lambda (n-2)}{n-1}\left( {{{\tilde{A}}}^{2n-4}}{{{\tilde{\Phi }}}^{2n-4}}-1 \right)} \right),
\end{aligned}
\end{equation}
and the above equation can be rewritten as
\begin{equation}
\tilde{Q}=-\frac{n-1}{4\Lambda }4\pi T\tilde{A}\tilde{\Phi }+\frac{n-1}{4\Lambda }\tilde{A}\tilde{\Phi }\sqrt{B-C{{{\tilde{\Phi }}}^{2n-4}}},
\end{equation}
where $B=16{{\pi }^{2}}{{T}^{2}}+\frac{8\Lambda (n-2)}{(n-1)}$, $C=\frac{8\Lambda (n-2)}{(n-1)}{{\tilde{A}}^{2n-4}}$.

Now again, considering a special case with $n=3$, Eq.(20) turns into
\begin{equation}
\begin{aligned}
{{Q}_{0}}({{\Phi }_{2}}-{{\Phi }_{1}})=&-\frac{\pi T}{\Lambda }\Phi _{2}^{2}-\frac{1}{3{{\Lambda }^{2}}}{{\left( 4{{\pi }^{2}}{{T}^{2}}+\Lambda -\Lambda \Phi _{2}^{2}\right)}^{\frac{3}{2}}}\\
 &+\frac{\pi T}{\Lambda }\Phi _{1}^{2}+\frac{1}{3{{\Lambda }^{2}}}{{\left( 4{{\pi }^{2}}{{T}^{2}}+\Lambda -\Lambda \Phi _{1}^{2}
 \right)}^{\frac{3}{2}}}.
\end{aligned}
\end{equation}
and
\begin{equation}
\begin{aligned}
&{{Q}_{0}}=(-\frac{2\pi T}{\Lambda }+\frac{1}{\Lambda }\sqrt{4{{\pi }^{2}}{{T}^{2}}+\Lambda -\Lambda \Phi _{2}^{2}}){{\Phi }_{2}},\\
&{{Q}_{0}}=(-\frac{2\pi T}{\Lambda }+\frac{1}{\Lambda }\sqrt{4{{\pi
}^{2}}{{T}^{2}}+\Lambda -\Lambda \Phi _{1}^{2}}){{\Phi }_{1}}.
\end{aligned}
\end{equation}
From the above Eq.~(24), one can easily obtain the following
expressions
\begin{equation}
\begin{aligned}
 0=&-2\pi T({{\Phi }_{2}}-{{\Phi }_{1}})\\
   &+{{\Phi }_{2}}\sqrt{4{{\pi }^{2}}{{T}^{2}}+\Lambda -\Lambda \Phi _{2}^{2}}-{{\Phi }_{1}}\sqrt{4{{\pi }^{2}}{{T}^{2}}+\Lambda -\Lambda \Phi _{1}^{2}},
\end{aligned}
\end{equation}
\begin{equation}
\begin{aligned}
 2{{Q}_{0}}=&-\frac{2\pi T}{\Lambda}({{\Phi }_{2}}+{{\Phi }_{1}})\\
            &+\frac{{{\Phi }_{2}}}{\Lambda }\sqrt{4{{\pi }^{2}}{{T}^{2}}+\Lambda -\Lambda \Phi _{2}^{2}}+\frac{{{\Phi }_{1}}}{\Lambda }\sqrt{4{{\pi }^{2}}{{T}^{2}}+\Lambda -\Lambda \Phi _{1}^{2}},
\end{aligned}
\end{equation}
Then substituting Eq.~(26) into Eq.~(23), we get
\begin{equation}
 \begin{aligned}
   &({{\Phi }_{2}}-{{\Phi }_{1}}) \left(-2\pi T({{\Phi }_{2}}+{{\Phi }_{1}})\right) \\
  &+({{\Phi }_{2}}-{{\Phi }_{1}})({{\Phi }_{2}}\sqrt{4{{\pi }^{2}}{{T}^{2}}+\Lambda -\Lambda \Phi _{2}^{2}})\\
  &+({{\Phi }_{2}}-{{\Phi }_{1}})({{\Phi }_{1}}\sqrt{4{{\pi }^{2}}{{T}^{2}}+\Lambda -\Lambda \Phi _{1}^{2}})\\
= &-2\pi T\Phi _{2}^{2}-\frac{2}{3\Lambda }{{\left( 4{{\pi }^{2}}{{T}^{2}}+\Lambda -\Lambda \Phi _{2}^{2} \right)}^{\frac{3}{2}}}\\
  &+2\pi T\Phi _{1}^{2}+\frac{2}{3\Lambda }{{\left( 4{{\pi }^{2}}{{T}^{2}}+\Lambda -\Lambda \Phi _{1}^{2} \right)}^{\frac{3}{2}}},
\end{aligned}
\end{equation}
If letting $x={{\Phi }_{1}}/{{\Phi }_{2}}$, $T=\chi {{T}_{c}}$, and
${{T}_{c}}=\frac{{{(-2\Lambda)}^{1/2}}}{3\pi }$ from Eq.~(15), the
combination of Eq.~(25) and (27) provides us
\begin{equation}\label{eq27}
0=-\frac{{{2}^{\frac{3}{2}}}}{3}\chi (1-x)+\sqrt{\frac{8}{9}{{\chi }^{2}}-1+\Phi _{2}^{2}}-x\sqrt{\frac{8}{9}{{\chi }^{2}}-1+{{x}^{2}}\Phi _{2}^{2}},
\end{equation}
\begin{equation}\label{eq28}
\begin{aligned}
    &(1-x)\left(\sqrt{\frac{8}{9}{{\chi }^{2}}-1+{{\Phi }_{2}}^{2}}+x\sqrt{\frac{8}{9}{{\chi }^{2}}-1+{{x}^{2}}{{\Phi }_{2}}^{2}} \right)\\
    &+(1-x)\left(-\frac{{{2}^{\frac{3}{2}}}}{3}\chi (1+x)\right) = -\frac{{{2}^{\frac{3}{2}}}}{3}\chi (1-{{x}^{2}})\\
    &-\frac{2}{3\Phi _{2}^{2}}{{\left( \frac{8}{9}{{\chi }^{2}}-1+\Phi _{2}^{2} \right)}^{\frac{3}{2}}}-\frac{2}{3\Phi _{2}^{2}}{{\left( \frac{8}{9}{{\chi }^{2}}-1+{{x}^{2}}\Phi _{2}^{2} \right)}^{\frac{3}{2}}},
 \end{aligned}
\end{equation}
From the above formulae, one can see that the value of $x$ and
$\Phi_2$ is independent of $\Lambda$. For a fixed $\chi$, i.e., a
fixed $T_0$, we can get a certain value for $x$ and $\Phi_2$ from
Eq.~(28) and (29).

Substituting Eq.~(24) into Eq.~(21), one can obtain similar formula
for high-dimensional space-time
\begin{equation}\label{eq30}
\begin{aligned}
  &\tilde{Q}=-\frac{n-1}{\sqrt{-2\Lambda }}\tilde{A}{{\tilde{\Phi }}_{2}}\\
  &\left( \frac{2(n-2)\chi }{(2n-3)}-\sqrt{\frac{4{{(n-2)}^{2}}{{\chi }^{2}}}{{{(2n-3)}^{2}}}+\frac{n-2}{n-1}\left( {{({\tilde{A}}}}\tilde{\Phi }_{2})^{2n-4}-1 \right)}
  \right).
\end{aligned}
\end{equation}
Note we can get the ${{\tilde{Q}}_{0}}$ in coexistence region from
the above equation. Fig.~5 shows the $Q-\Phi$ line on the background
of isotherms at different temperature. When the temperature is lower
than $T_c$, the isopotential line will replace the curve which does
not meet the requirements of thermodynamic stability. The numerical
values of $\chi$, $x$, $\Phi_1$, $\Phi_2$, $T_0$, and $Q_0$ at
different space-time dimensions are also explicitly illustrated in
Table 1.

\begin{figure*}[htbp]
  \begin{center}
   \resizebox{0.4\textwidth}{!}{\includegraphics{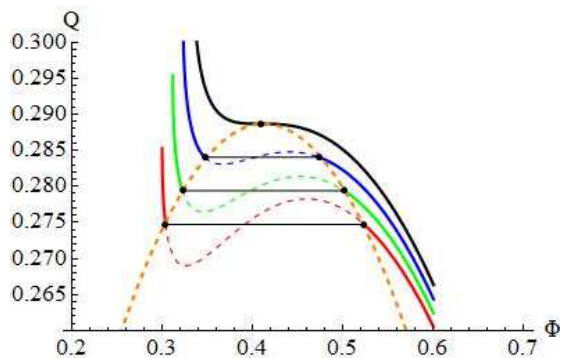}}
\caption{The simulated isothermal phase transition by isobars and
the boundary of two-phase coexistence region for RN-Ads black hole.
The boundary of the two-phase equilibrium region is denoted by
dotted dashed curve with $n=3$, and $\chi=1 (black)$, $\chi=1.004
(blue)$, $\chi=1.008 (green)$, $\chi=1.012 (red)$. The area enclosed
by yellow dotted lines represents the two-phase coexistence region.
}
  \end{center}
\end{figure*}

\begin{table}[h]
\caption{State quantities at phase transition points, considering
different space-time dimensions ($\Lambda=-1$).}
\label{tab:1}       
\scriptsize
\begin{tabular}{lllllll}
\hline\noalign{\smallskip}
$n$ & $\chi$ & $x$ & $\Phi_1$ & $\Phi_2$ & $Q_0$ & $T_0$ \\
\noalign{\smallskip}\hline\noalign{\smallskip}
 & 1 & 1 & 0.408248 & 0.408248 & 0.288675 & 0.150053\\
 & 1.002 & 0.802808 & 0.364321 & 0.453809 & 0.286363 & 0.150353\\
3& 1.004 & 0.732417 & 0.346572 & 0.473190 & 0.284047 & 0.150653\\
 & 1.006 & 0.682245 & 0.333122 & 0.488273 & 0.281726 & 0.150953\\
\noalign{\smallskip}\hline
 & 1 & 1 & 0.608376 & 0.608376 & 0.516398 & 0.180063\\
 & 1.002 & 0.643326 & 0.482451 & 0.749933 & 0.503992 & 0.180423\\
4& 1.004 & 0.533619 & 0.434306 & 0.813887 & 0.491562 & 0.180784\\
 & 1.006 & 0.460886 & 0.398823 & 0.865339 & 0.479109 & 0.181144\\
\noalign{\smallskip}\hline
 & 1 & 1 & 0.919546 & 0.919546 & 0.540560 & 0.200070\\
6& 1.001 & 0.534718 & 0.660438 & 1.235115 & 9.159250 & 0.200270\\
 & 1.002 & 0.408734 & 0.566763 & 1.386630 & 8.778500 & 0.200470\\
\noalign{\smallskip}\hline
\end{tabular}
\end{table}

In the canonical ensemble with fixed charge, the potential, which is
also the free energy of the system, presents the thermodynamic
behavior of a system in a standard approach. However, in our
analysis we will consider an extended phase space. According to
first law of black hole thermodynamics and the interpretation of $M$
(total mass of black hole) \cite{35,56} as $H$ (the black hole
enthalpy) [55,56], the Gibbs free energy  of black hole can be
written as
\begin{equation}\label{eq30}
\begin{aligned}
G =&M-TS\\
  =&\frac{1}{8\Gamma \left(\frac{n}{2} \right)}{{\pi }^{\frac{n}{2}-1}}{{r_+}^{n-2}}\left\{-Q^{\frac{1}{2-n}} r_+ {\left( Q{{r_+}^{2-n}} \right)}^{\frac{1}{2-n}} \right.\\
  &\left.\left(\left(n-2\right)\left({{\left(Q{{r_+}^{2-n}}\right)}^{\frac{2}{n-2}}}-{{\left(Q{{r_+}^{2-n}}\right)}^{\frac{2n-2}{n-2}}}\right)\right.\right.\\
 &\left.\left. -\frac{2{{Q}^{\frac{2}{n-2}}}\Lambda }{n-1} \right)+\left(n-1 \right)\left( 1+{{Q}^{2}}{{r_+}^{4-2n}}+\frac{2{{r_+}^{2}}\Lambda }{n-{{n}^{2}}} \right)\right\},
 \end{aligned}
\end{equation}
Here, according to Eq.~(9), $r_+$ is a function of charge and
temperature, $r_+=r_+(Q,T)$. In Fig.~6-7, we plot the change of the
free energy $G$ with $T$ (for fixed $Q$) and Q (for fixed $T$) in
different space-time dimensions. The existence of "swallow tail"
behavior is clearly revealed, which indicates that the small-large
black hole phase transition occurring in the system is of the first
order.

\begin{figure*}[htbp]
  \begin{center}
    \mbox{
      \subfigure[$n=3$]{\scalebox{0.4}{\includegraphics{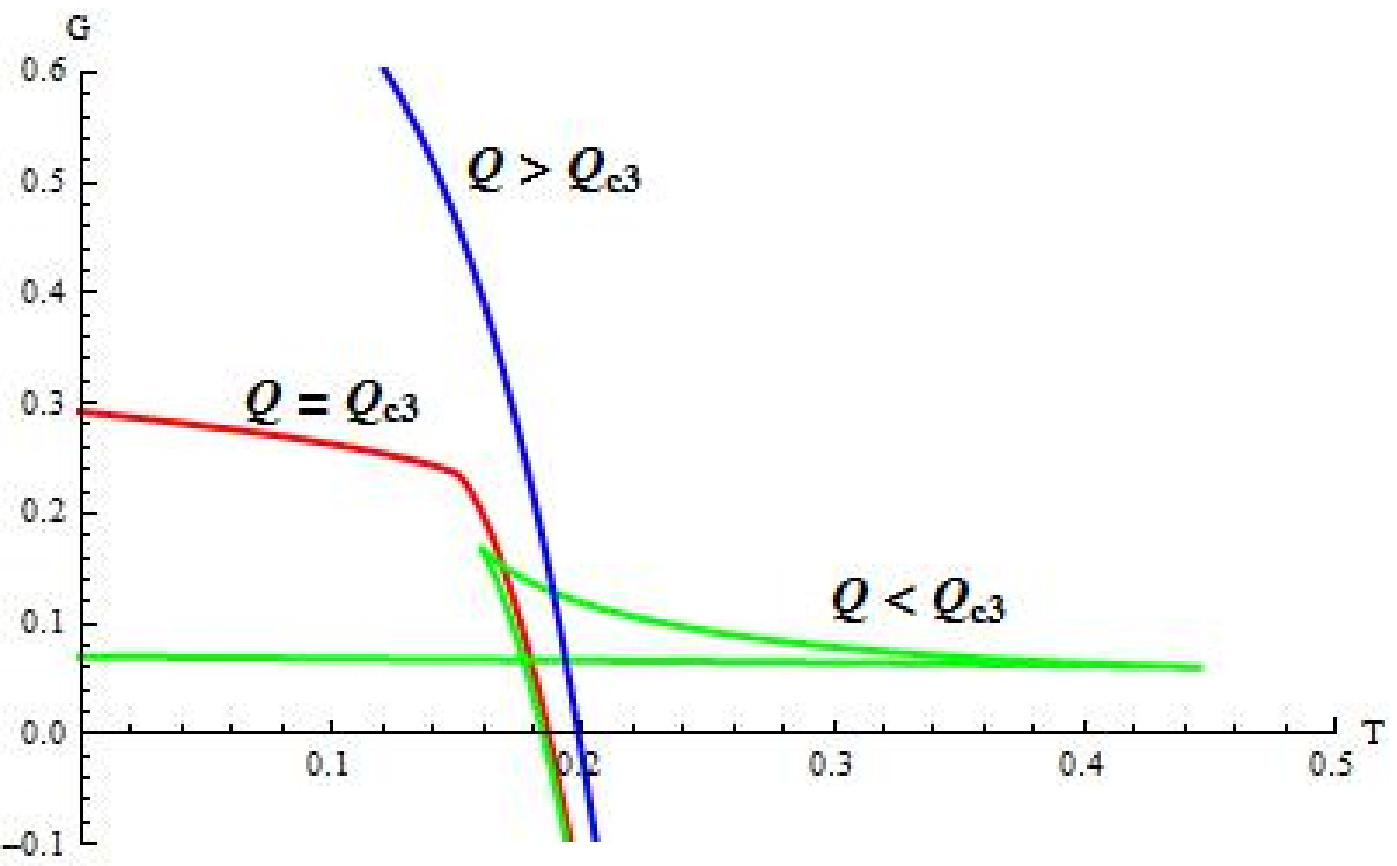}}} \quad
      \subfigure[$n=4$]{\scalebox{0.4}{\includegraphics{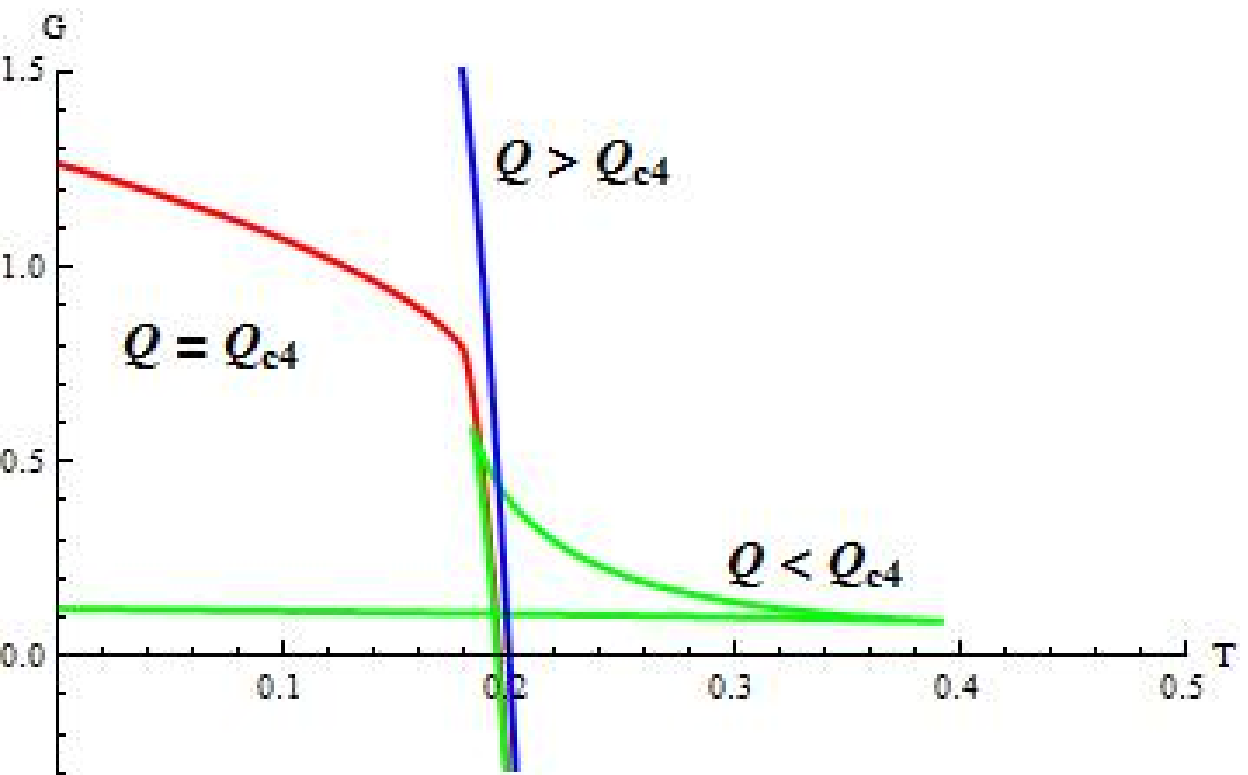}}}
      }
       \caption{Gibbs free energy versus $T$ for (a) $n=3, \Lambda=-1$ and (b) $n=4,
       \Lambda=-1$.
       }
  \end{center}
\end{figure*}

\begin{figure*}[htbp]
  \begin{center}
    \mbox{
      \subfigure[$n=3$]{\scalebox{0.4}{\includegraphics{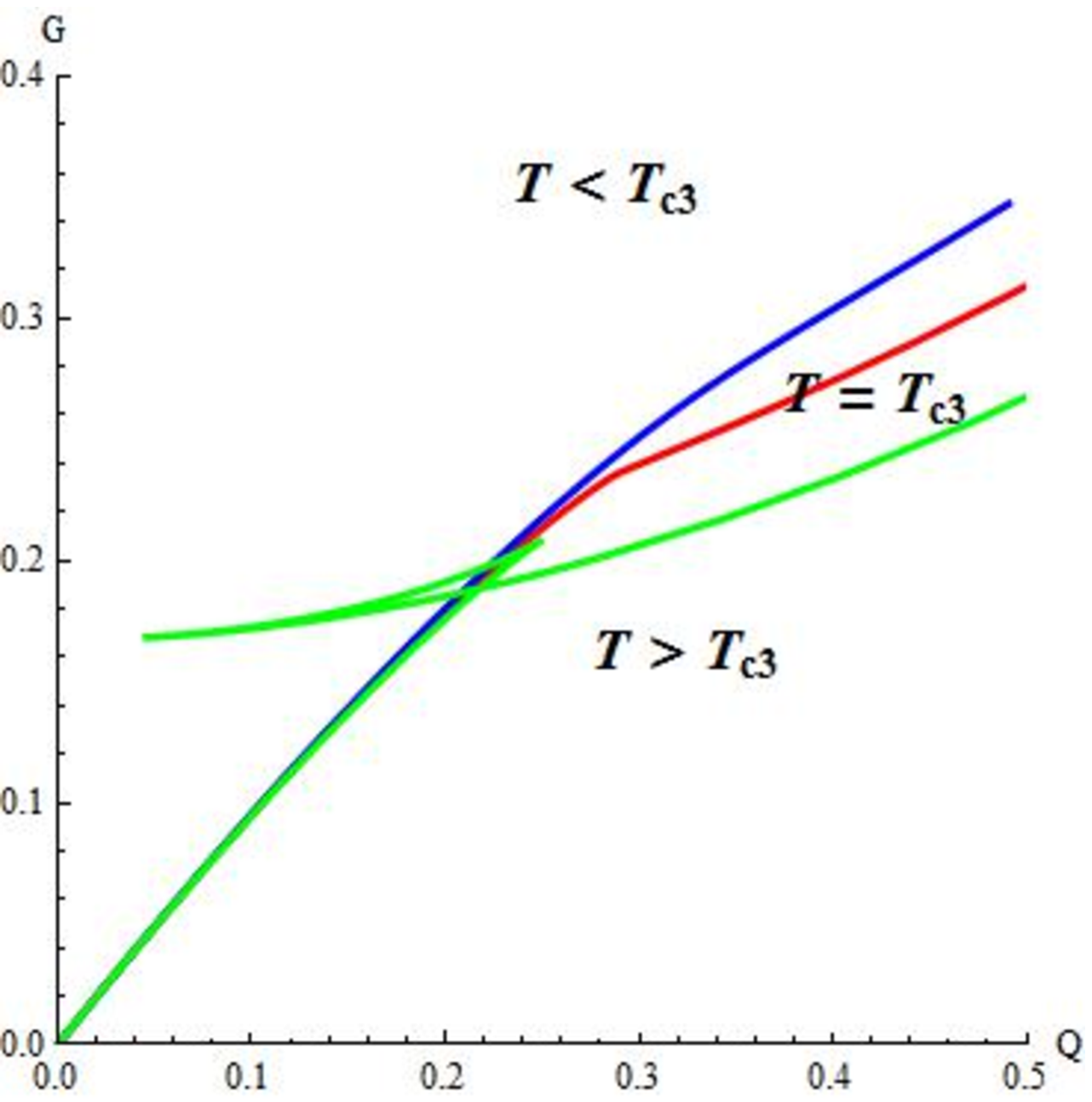}}} \quad
      \subfigure[$n=4$]{\scalebox{0.4}{\includegraphics{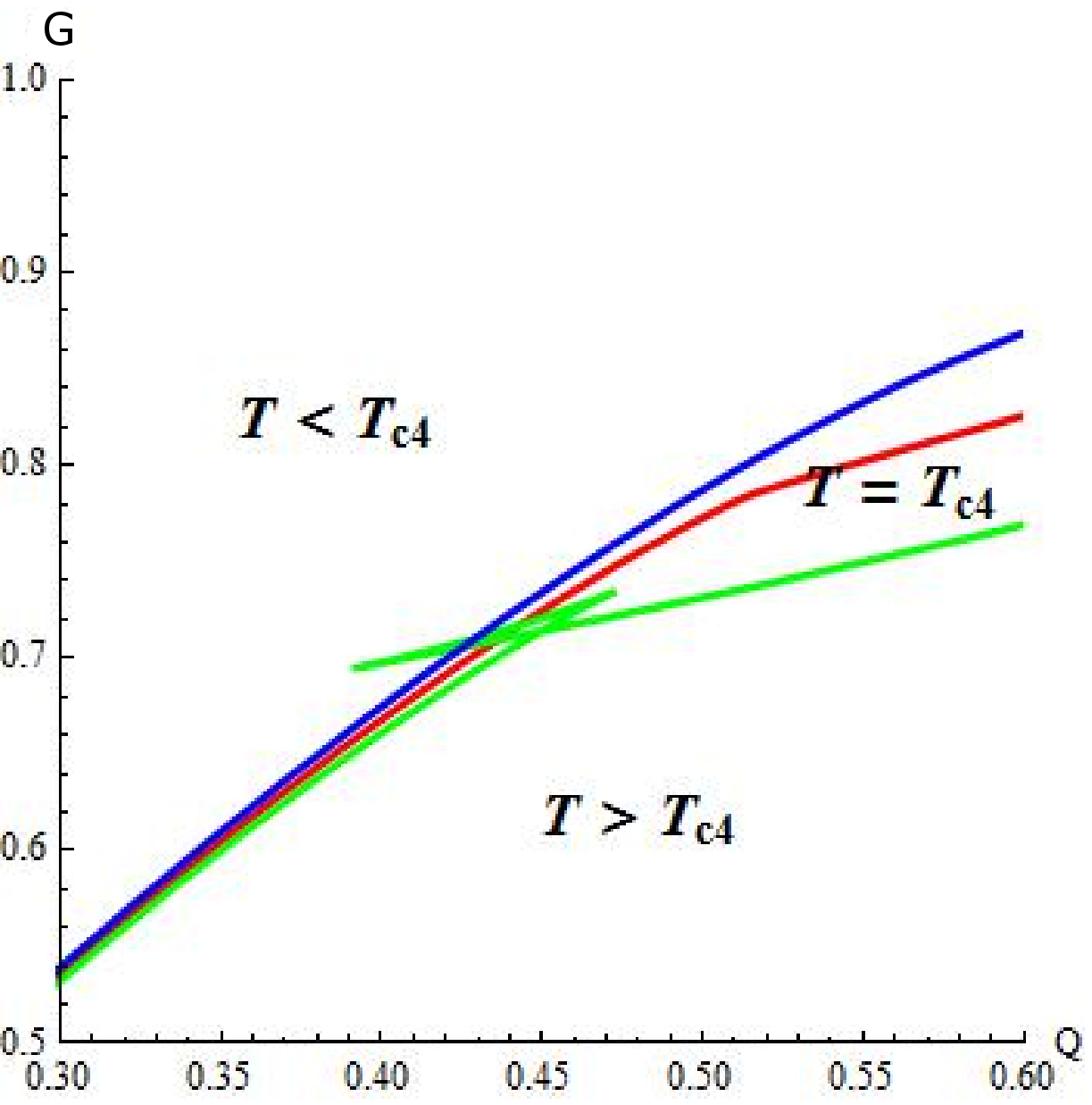}}}
      }
       \caption{Gibbs free energy versus $Q$ for (a) $n=3, \Lambda=-1$ and (b) $n=4,
       \Lambda=-1$.
       }
  \end{center}
\end{figure*}

And in Fig.8 and Fig.9, we also plot the $G-T$ curves at the same dimension and different values of $\Lambda$. It is shown that the $G-T$ criticality nearly unchanges. Only the position of the critical point changes.

\begin{figure*}[htbp]
  \begin{center}
    \mbox{
      \subfigure[$n=3,\Lambda=-0.5$]{\scalebox{0.5}{\includegraphics{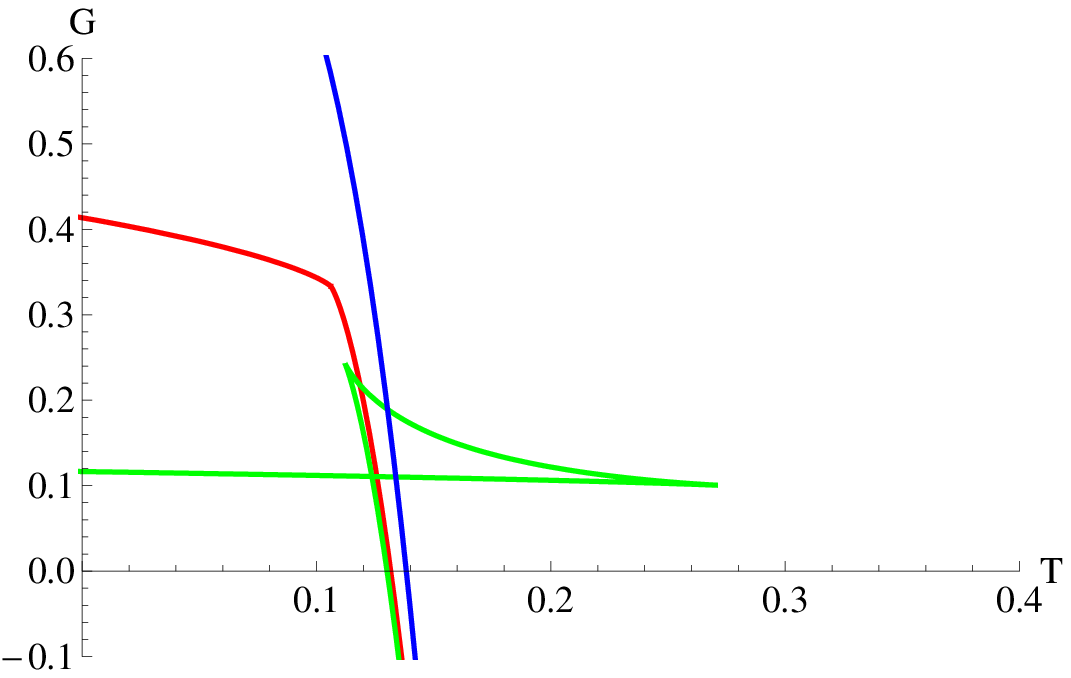}}} \quad
      \subfigure[$n=3,\Lambda=-1$]{\scalebox{0.5}{\includegraphics{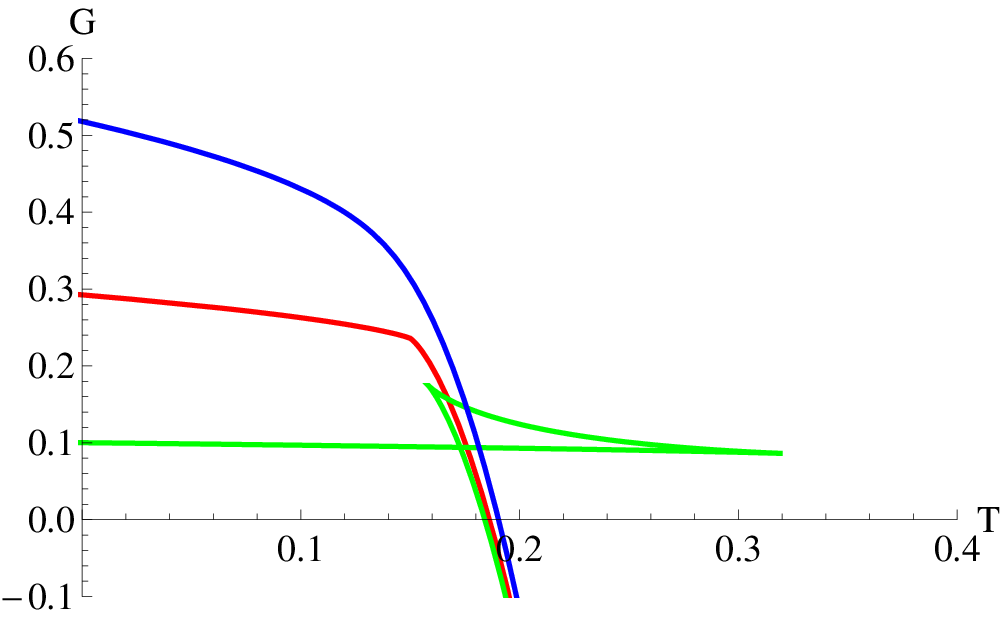}}}
      }
    \mbox{
      \subfigure[$n=3,\Lambda=-2$]{\scalebox{0.5}{\includegraphics{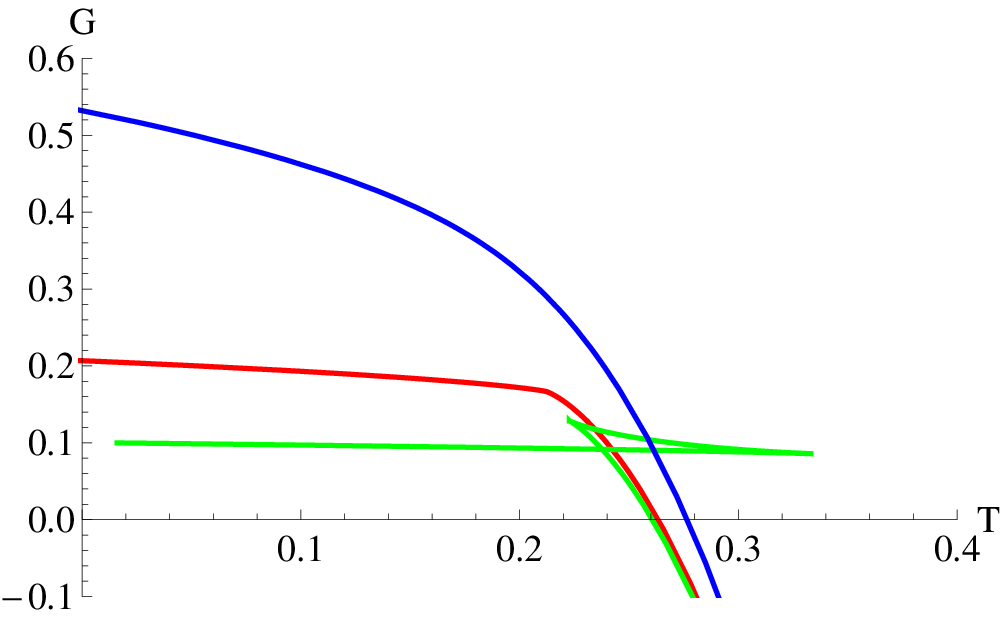}}} \quad
      \subfigure[$n=3,\Lambda=-3$]{\scalebox{0.5}{\includegraphics{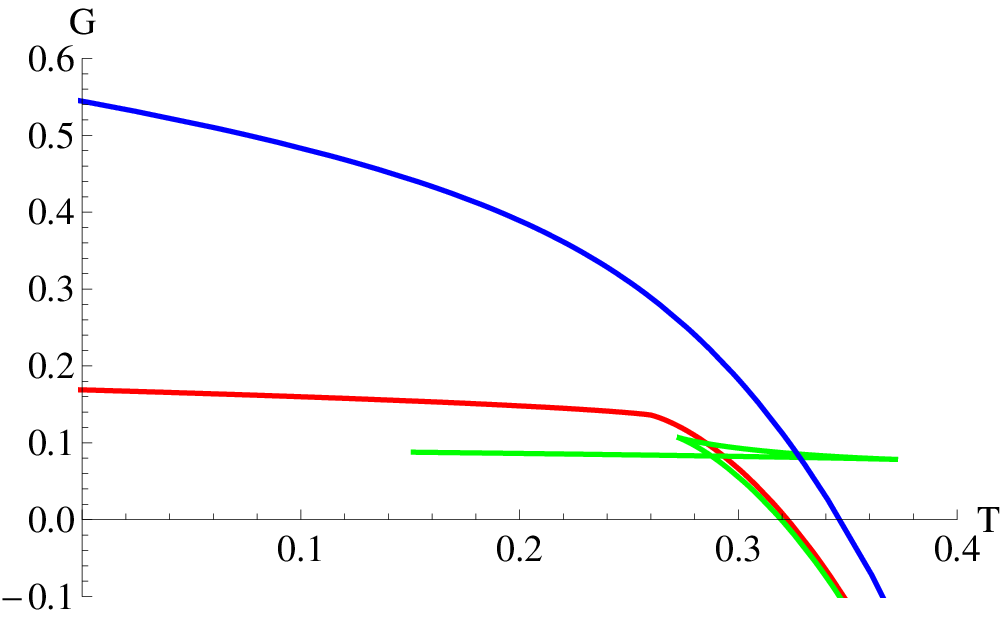}}}
      }
    \caption{Gibbs free energy versus $T$ for $n=3$, and $\Lambda=-0.5,-1,-2,-3$, respectively.
    }
    \label{Qn}
  \end{center}
\end{figure*}

\begin{figure*}[htbp]
  \begin{center}
   \resizebox{0.4\textwidth}{!}{\includegraphics{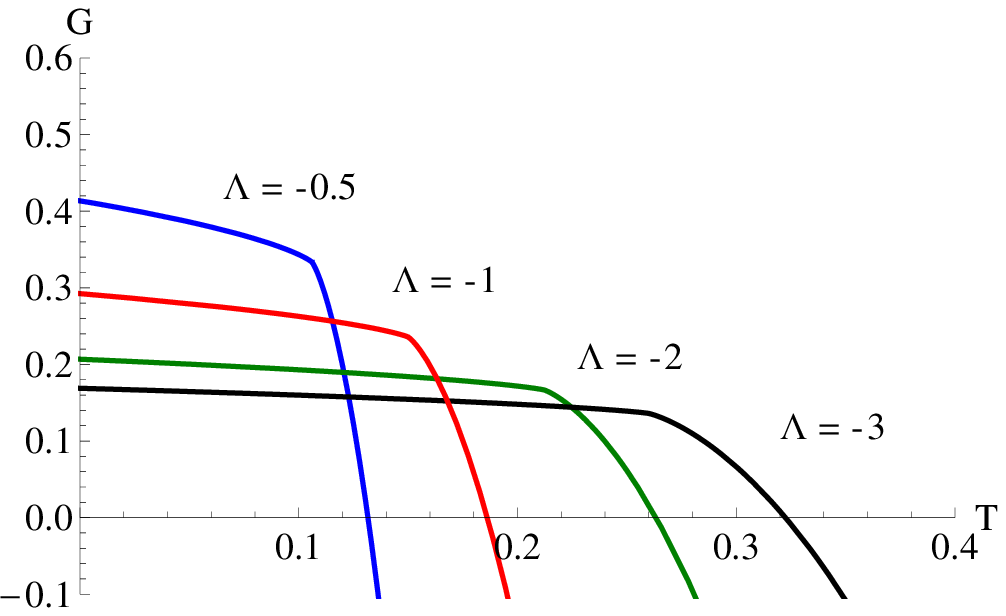}}
\caption{Gibbs free energy versus $T=Tc$ for $n=3$, and $\Lambda=-0.5,-1,-2,-3$, respectively.
}
  \end{center}
\end{figure*}

From Fig.~6-7, we find that general $(n+1)$-dimensional RN-AdS black
holes thermodynamic systems have typical characteristics of Van's
system gas/liquid phase transition. If we do not treat the cosmological constant as a thermodynamic variable and consider the non-extended phase space, black hole mass $M$ now should be the internal energy of the system , the Gibbs free energy is defined as the following form [29]
\begin{equation}\label{eq32}
\begin{aligned}
 G=&M-TS-Q\Phi\\
  =&\frac{\left( n-1\right){{\pi}^{\frac{n}{2}-1}{{r_+}^{n-2}}\left( 1+{Q^2}{{r_+}^{4-2n}}-\frac{2{{r_+}^2}\Lambda}{\left( n-1\right){n}} \right)}}{8\Gamma (\frac{n}{2})}\\
  &-\frac{1}{8\Gamma (\frac{n}{2})}{{\pi}^{\frac{n}{2}-1}}{Q^{-\frac{1}{n-2}}}{{{r_+}}^{n-1}}{{\left( Q{{r_+}}^{2-n} \right)}^{-\frac{1}{n-2}}}\\
  &\left( \left( n-2 \right){{\left( Q{{{r_+}}^{2-n}} \right)}^{\frac{2}{n-2}}}-\left( n-2 \right){{\left( {{\left( Q{{{r_+}}^{2-n}} \right)}^{\frac{2n-2}{n-2}}}\right)}}\right.\\
  &\left.-\frac{2{{Q}^{\frac{2}{n-2}}}\Lambda}{n-1}\right)-\frac{\left( n-1\right){{\pi^{\frac{{n}}{2}-1}}{Q^2}{{r_+}^{2-n}}}}{4\Gamma (\frac{n}{2})},
\end{aligned}
\end{equation}
In Fig.~10-11 we plot the change of the free energy $G$ with $T$ and
$Q$, for fixed $\Lambda=-1$ and different space-time dimensions.
Fig.~10 reveals the existence of "swallow tail" behavior of the free
energy. However, due to a distinct definition for Gibbs free energy,
we fail to detect the "swallow tail" in Fig.~11, which is different
from the case shown in Fig.~7.

\begin{figure*}[htbp]
  \begin{center}
    \mbox{
      \subfigure[$n=3$]{\scalebox{0.4}{\includegraphics{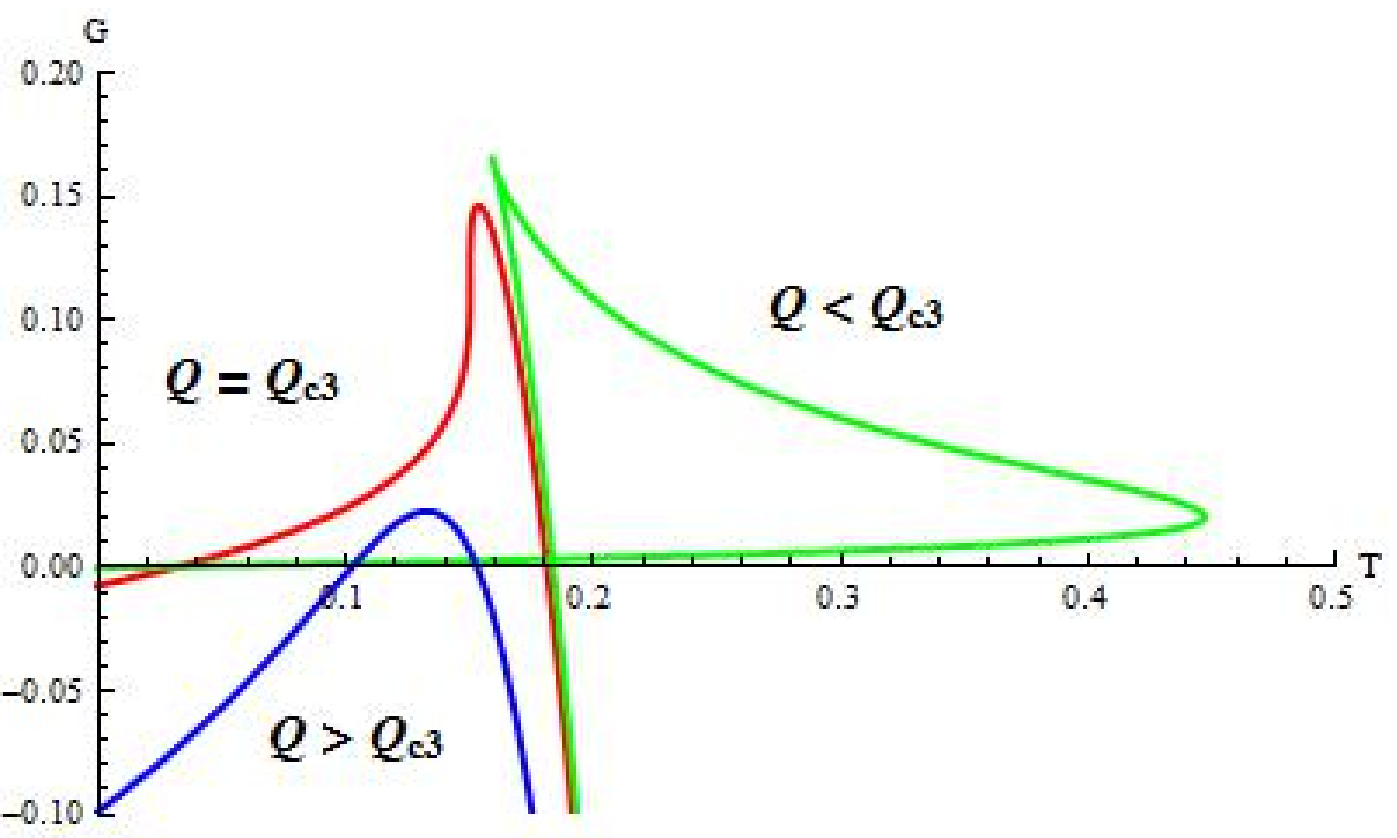}}} \quad
      \subfigure[$n=4$]{\scalebox{0.4}{\includegraphics{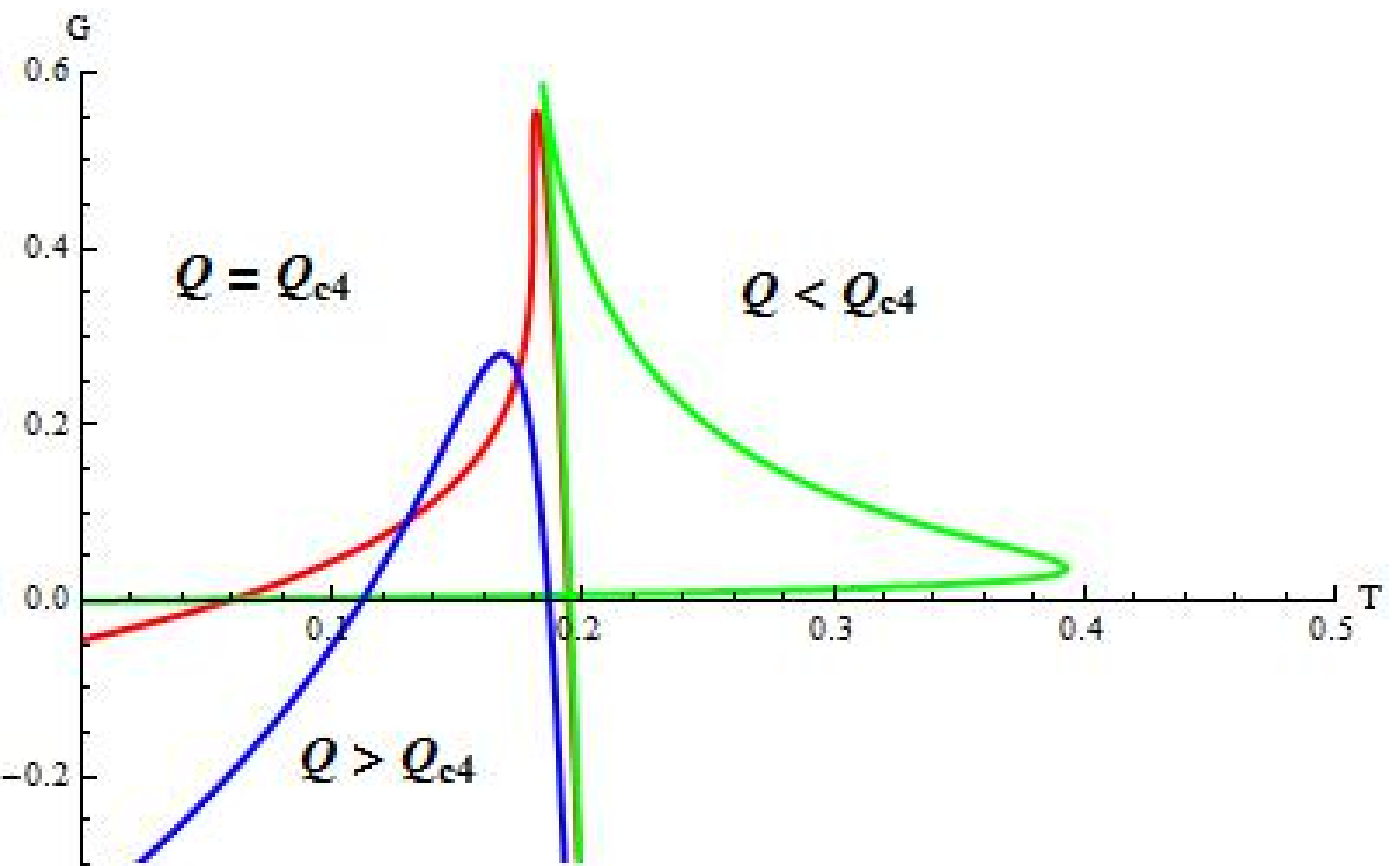}}}
      }
       \caption{Gibbs free energy versus T for (a) $n=3, \Lambda=-1$ and for (b) $n=4, \Lambda=-1$
       }
  \end{center}
\end{figure*}

\begin{figure*}[htbp]
  \begin{center}
    \mbox{
      \subfigure[$n=3$]{\scalebox{0.45}{\includegraphics{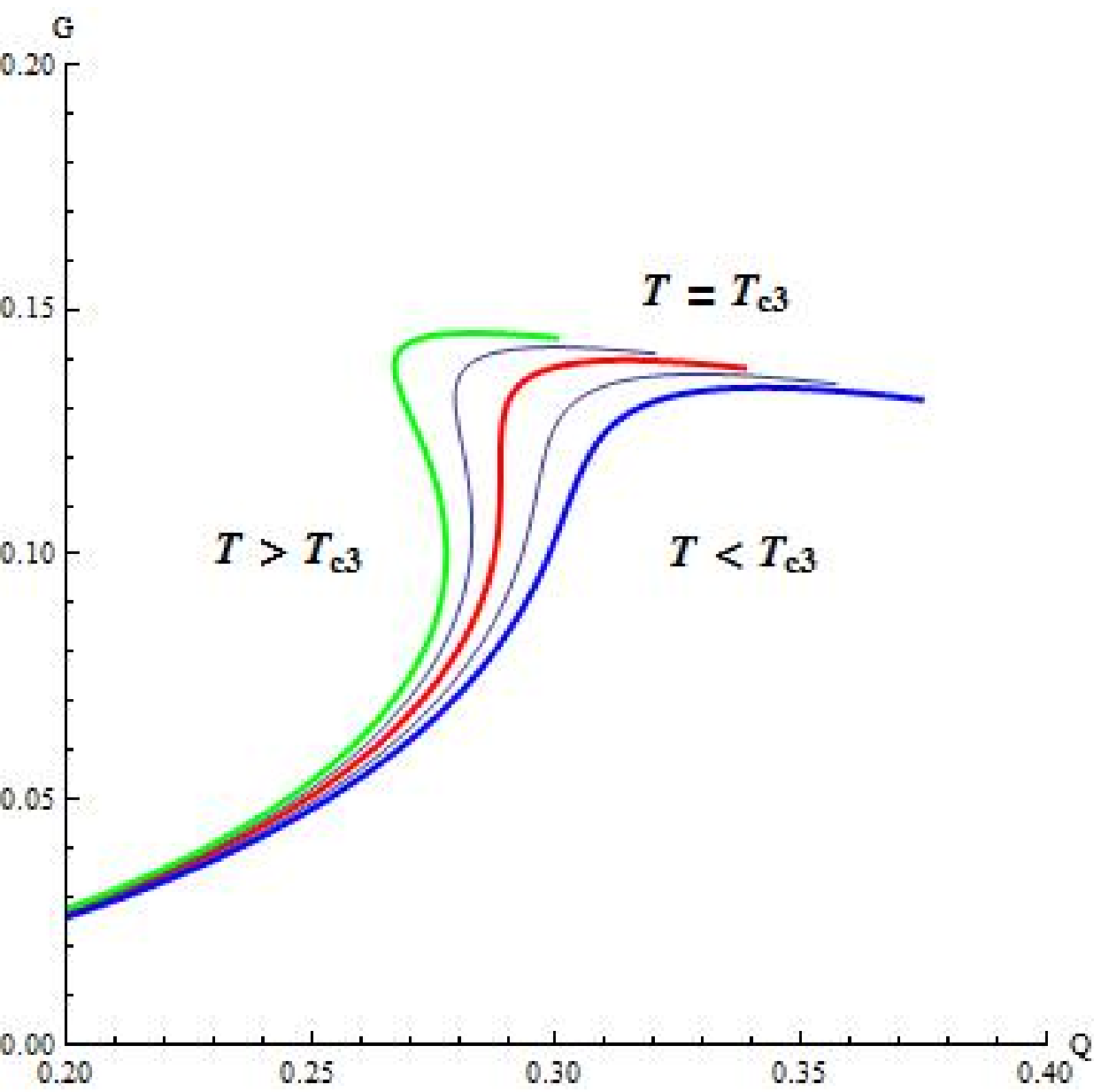}}} \quad
      \subfigure[$n=4$]{\scalebox{0.45}{\includegraphics{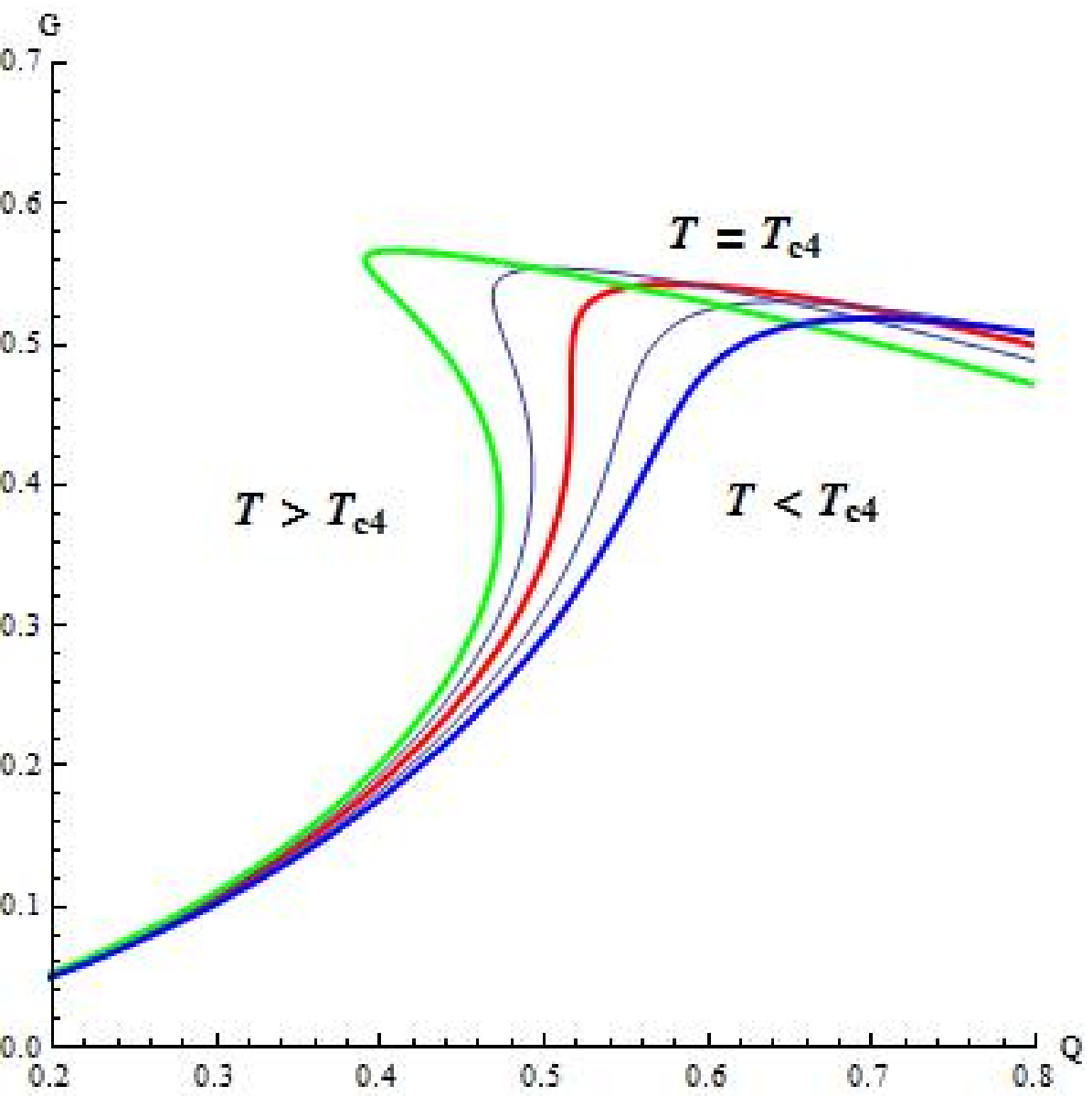}}}
      }
       \caption{Gibbs free energy versus $Q$ for (a) $n=3, \Lambda=-1$ and for (b) $n=4, \Lambda=-1$
       }
  \end{center}
\end{figure*}

\section{Discussion} \label{sec:3}

Taking general $(n+1)$-dimensional RN-AdS black holes as
thermodynamic systems, the state equation of which is meaningless in
some region. Using Maxwell's equal area law (deduced from minimum
Gibbs free energy theory) and taking phase transition into
consideration, the meaningless region in the state equation no
longer exists. Fig.~1 and Fig.~5 show that, when the system is at
constant temperatures higher than the critical temperature, the
$Q-\Phi$ curves are partially replaced by the isotherm and
isopotential lines, which implies $Q$ and $T$ are invariants while
potential $\Phi$ is changing. This region is a two-state coexistence
region, where the phase transition is of first order according to
Ehrenfest classification.

From the discussion above, we know that for general
$(n+1)$-dimensional RN-AdS black holes thermodynamic system, when
taking $(T,Q,\Phi )$ as state parameters, the system shows similar
phase transition characteristics to that of Van's system. The
position of the critical point is also the same as the case when
taking $(T,P,V)$ as state parameters.

Moreover, by applying Maxwell's equal area law to phase transition
behaviors of thermodynamic system, we have derived both the position
of critical point and the two phase coexistence region, which make
it possible to obtain a more clear understanding of the phase
transition process of such systems \cite{32}.

Taking Ads black hole as a thermodynamic system, it was found that
the phase transition of various Ads black holes are similar to that
of the Vander waals-Maxwell gas liquid \cite{1,34,36}. Therefore, we
can find some observable systems (Vander waals gas) similar to the
Ads and ds background black holes. Considering the similarities they
share in the thermodynamic properties, we may work backward and
investigate other properties of black holes, such as phase
transition and critical behaviors. This study will further
contribute to a deeper understanding of black hole entropy,
temperature, and thermal capacity, as well as the completion of
self-consistent black hole thermodynamics.


\section*{Acknowledgments}

This work was supported by the Ministry of Science and Technology
National Basic Science Program (Project 973) under Grants Nos.
2012CB821804 and 2014CB845806, the Strategic Priority Research
Program ``The Emergence of Cosmological Structure" of the Chinese
Academy of Sciences (No. XDB09000000), the National Natural Science
Foundation of China under Grants Nos. 11475108, 11503001, 11373014
and 11073005, the Fundamental Research Funds for the Central
Universities and Scientific Research Foundation of Beijing Normal
University, China Postdoctoral Science Foundation under grant No.
2015T8 0052, and the Opening Project of Key Laboratory of
Computational Astrophysics, National Astronomical Observatories,
Chinese Academy of Sciences.

%

\end{document}